\documentclass[a4paper,10pt,twoside]{cpc-hepnp}
\usepackage{multicol}
\usepackage{graphicx}
\usepackage{booktabs}
\usepackage{multirow}
\usepackage{amssymb,bm,mathrsfs,bbm,amscd}
\usepackage[tbtags]{amsmath}
\usepackage{lastpage}
\usepackage{CJK}
\usepackage{epstopdf}
\usepackage{epsfig}
\usepackage{color}
\usepackage{ifpdf}
\usepackage{hyperref}
\usepackage[thicklines]{cancel}

\begin{document}

\title{Density-dependent relativistic mean field approach and its application to single-$\Lambda$ hypernuclei in Oxygen isotopes \thanks{This work was partly supported by the Fundamental Research Funds for the Central Universities, Lanzhou University under Grant No. lzujbky-2022-sp02 and lzujbky-2023-stlt01, the National Natural Science Foundation of China under Grant No. 11875152 and No. 12275111, and the Strategic Priority Research Program of Chinese Academy of Sciences under Grant No. XDB34000000. The authors also want to thank the computation resources provided by the Supercomputing Center of Lanzhou University.}}
\author{
      Shi-Yuan Ding$^{1,2)}$
\quad Wei Yang$^{1,2}$
\quad Bao-Yuan Sun$^{1,2;1)}$\email{sunby@lzu.edu.cn}
}
\maketitle
\address{
 $^1$MOE Frontiers Science Center for Rare Isotopes, Lanzhou University, Lanzhou 730000, China\\
 $^2$School of Nuclear Science and Technology, Lanzhou University, Lanzhou 730000, China
}

\begin{abstract}
  The in-medium feature of nuclear force which includes both nucleon-nucleon ($NN$) and hyperon-nucleon ($\Lambda N$) interactions impacts the description of single-$\Lambda$ hypernuclei. With the alternated mass number or isospin of hypernuclei, such effects could be unveiled by analyzing systematical evolution of the bulk and single-particle properties. From a density-dependent meson-nucleon/hyperon coupling perspective, a new $\Lambda N$ effective interaction in the covariant density functional (CDF) theory, namely DD-LZ1-$\Lambda1$, is obtained by fitting the experimental data of $\Lambda$ separation energies for several single-$\Lambda$ hypernuclei. It is then adopted to study the structure and transition properties of single-$\Lambda$ hypernuclei in Oxygen isotopes, comparing with several selected CDF Lagrangians. Discrepancy is observed explicitly in the isospin evolution of $\Lambda1p$ spin-orbit splitting with various effective interactions, ascribed to their divergence of the meson-hyperon coupling strengths with increasing density. In particular, the density-dependent CDFs introduce an extra contribution to enhance the isospin dependence of the splitting, which is originated from the rearrangement terms of $\Lambda$ self-energies. In addition, the characteristics of hypernuclear radii are studied along the isotopic chain. Owing to the impurity effect of $\Lambda$ hyperon, a size shrinkage is observed in the matter radii of hypernuclei as compared to their cores of normal nuclei, while its magnitude is elucidated further to correlate with the incompressibility of nuclear matter. Besides, there exists a sizable model-dependent trend that $\Lambda$ hyperon radii evolve with the neutron number, which is decided partly by the in-medium $NN$ interactions as well as the core polarization effects.
\end{abstract}


\section{Introduction}\label{Introduction}

The discovery of hyperon, particles containing strange quarks, in 1953 sparked strong interest among experimental and theoretical physicists \cite{Danysz1953Philos.Mag.44.348}. The ability of hyperons to enter the nucleus and form a system of hypernuclei makes them sensitive probes for studying the structure and specific nuclear features. The studies on hyperon behavior in the nucleus help us to understand the baryon-baryon interaction in nuclear medium and its effects on nuclear properties \cite{Hashimoto2006PPNP57.564, Gal2016Rev.Mod.Phys.88.035004}. In addition, hyperons are thought to be produced inside neutron stars \cite{Prakash9971Phys.Rep.280.1, Tolos2020PPNP112.103770, Burgio2021PPNP120.103879}. The link between hypernucleus and neutron star properties benefits our comprehension of the state of matter in extreme environments, as well as strangeness-bearing nuclear force at high densities. In recent decades, a wealth of hypernuclear data has been generated through induced reactions of meson and electron beams at various radioactive beam facilities, including the Japan Proton Accelerator Research Complex (J-PARC) \cite{Sawada2007NPA782.434}, the Thomas Jefferson National Accelerator Facility (JLab) \cite{Nakamura2005NPA754.421}, and the Facility for Antiproton and Ion Research (FAIR) \cite{Henning2004NPA734.654}. These advanced facilities have played a pivotal role in advancing our understanding of strangeness in nuclear physics. Notably, single-$ \Lambda $ hypernuclei have been the most extensively studied, with experimental data covering hypernuclei from $^{3}_\Lambda$H to $^{208}_\Lambda$Pb in various laboratories \cite{Pile1991PRL66.2585, Hashimoto2006PPNP57.564, Feliciello2015Rep.Prog.Phys.78.096301, Gal2016Rev.Mod.Phys.88.035004}.

When $\Lambda$ hyperon enters into a nucleus, various phenomena could be observed. For instance, in $^{7}_{\Lambda}$Li, it has been found that the size of the $^{6}$Li core is smaller compared to the free space $^{6}$Li nucleus, as suggested by the measurement of the $ \gamma $-ray transition probability from $ E2(5/2^{+}\rightarrow1/2^{+}$) in $^{7}_{\Lambda}$Li \cite{Tanida2001PRL86.1982}. In addition, in $^{13}_{\Lambda}$C, it is hinted that the $\Lambda$ spin-orbit splitting is much smaller than the nucleon's \cite{Kohri2002PRC65.034607}. Recently, the potential for producing neutron-rich hyperfragments at high-intensity heavy-ion accelerator facilities is discussed \cite{Feng2020PRC102.044604, Saito2021Nat.Rev.Phys.3.803}. The directed flow of hypernuclei ($ ^{3}_{\Lambda} $H and $ ^{4}_{\Lambda} $H) just observed at RHIC for the first time in heavy-ion collisions, providing insights into hyperon-nucleon interactions under finite pressure \cite{Aboona2023PRL130.212301}. These advances highlight the promising prospects for investigating hypernuclear structures using the forthcoming high-intensity heavy-ion accelerator facility HIAF \cite{Yang2013NIMPR317.263, Zhou2022AAPPSBulletin32.35}. To provide accurate predictions for these experiments, researchers have performed detailed theoretical work on observables such as hypernuclear binding energy \cite{Mares1994PRC49.2472, Wirth2018PLB779.336}, spin-orbit splitting \cite{Vretenar1998PRC57.R1060, Umeya2011PRC83.034310, Xia2017Sci.China-Phys.Mech.Astron60.102021}, hyperon and hypernuclear matter radius \cite{Wei2009CPC33.116, Lu2011PRC84.014328, Zhang2021PRC103.034321, Zhang2022PTEP2022.023D01, Xue2022PRC106.044306}. Overall, these efforts aim to provide valuable insights into the behavior of hypernuclei, and to deepen our understanding of the in-medium baryon interactions.

Due to their ability to provide a self-consistent and unified description of almost all nuclei on the nuclear chart, both non-relativistic and relativistic mean-field theories are widely used in the calculation of finite nuclei and nuclear matter, and have been extended to describe hypernuclear systems with strange degrees of freedom during the development of theoretical models \cite{Reinhard1989Rep.Prog.Phys.52.439, Ring1996PPNP37.193, Bender2003Rev.Mod.Phys.75.121, Vretenar2005Phys.Rep.409.101, Meng2006PPNP57.470, Niksic2011PPNP66.519, Meng2015JPG42.093101, Meng2016Density, Rayet1976Ann.Phys.102.226, Lanskoy1997PRC55.2330, Brockmann1977PLB69.167, Bouyssy1981PLB99.305, Glendenning1991PRL67.2414, Mares1994PRC49.2472, Sugahara1994PTP92.803, Vretenar1998PRC57.R1060, Zhou2008PRC78.054306, Hu2014PRC90.014309, Li2018EPJA54.133, Zhou2020PLB807.135533}. As a key model utilized in this work, the relativistic mean-field theory has been extensively developed to study hypernulcear properties such as hyperon separation energy \cite{Mares1994PRC49.2472, Yao2017PRC95.034309}, spin-orbit splitting \cite{Brockmann1977PLB69.167, Tanimura2012PRC85.014306, Xia2017Sci.China-Phys.Mech.Astron60.102021}, hyperon halo \cite{Lv2002CPL19.1775}, hypernuclear deformation \cite{Win2008PRC78.054311, Lu2011PRC84.014328, Zhou2014PRC89.044307, Chen2022CPC46.064109}, cluster structure \cite{Tanimura2019PRC99.034324} and drip lines \cite{Meng2003NPA722.C366}. While most theoretical models have primarily emphasized nonlinear self-coupling interactions for studying hypernuclei, there has been a recent study that explores the effective interactions for single-$ \Lambda $ hypernuclei within the density-dependent relativistic mean-field (DDRMF) model \cite{Rong2021PRC104.054321}. With three distinct fitting approaches, they propose six new sets of effective $ \Lambda N $ interactions and uncover a significant linear correlation between the ratios $ R_{\sigma} $ and $ R_{\omega} $, representing scalar and vector coupling strengths, respectively, between these effective $ \Lambda N $ and $ NN $ interactions.

Recently, a new type of density-dependent relativistic mean-field Lagrangian, DD-LZ1, has been proposed, inspired by the restoration of pseudo-spin symmetry (PSS) and nuclear medium effects \cite{Wei2020CPC44.074107}. This new effective Lagrangian has produced satisfactory results in describing the properties of nuclear matter and finite nuclei. With unique density-dependent form, DD-LZ1 eliminates the spurious shell closures that appeared in previous RMF calculations, and reasonably restores the PSS of high orbital angular momentum near the Fermi energy \cite{Wei2020CPC44.074107}. Applications with this new RMF Lagrangian has been performed for several nuclear many-body characteristics, in both finite nuclei with mass ranging from light to superheavy, and neutron star properties with density ranging from low to high. For instance, a comprehensive macroscopic-microscopic model was developed to evaluate the total energies for even-even nuclei with proton numbers ranging from 8 to 110 \cite{Zhang2022CPC46.104.105}. Even with the appearance of hyperon \cite{Rather2021APJ917.46, Sun2023APJ942.55}, larger maximum masses of neutron stars could be obtained with DD-LZ1 than with several other RMF parameter sets, providing the possibility that the secondary object observed in GW190814 is a neutron star \cite{Rather2021PRC103.055814, Malik2022APJ930.17, Yang2022PRD105.063023}. Utilizing the Thomas-Fermi approximation, different microscopic structures of nonuniform nuclear matter were calculated for the crust of neutron stars and a unified equation of state was established in a vast density range \cite{Xia2022PRC105.045803, Xia2022CTP74.095303}. The different density-dependent behaviors of meson-nucleon couplings impact the microscopic structures of neutron star matter with DD-LZ1, affect correspondingly the description on various physical processes and evolutions of neutron stars.

Apart from dealing with the different nuclear medium effects caused by the interactions themselves, the evolution of isospin also leads to significant changes in the in-medium effects of hypernuclei, thereby affecting the description of their structural properties. In recent years, a series of refined theoretical studies have been conducted on hypernuclei in different isotopic chains using various interaction models. For instance, the no-core shell model has been employed to investigate the systematic evolution of the ground and excited state energies in the Helium and Lithium hyperisotopes \cite{Wirth2018PLB779.336}. The antisymmetrized molecular dynamics method has been applied to explore the e low-lying level structure of hypernuclei in the Beryllium hyperisotopes \cite{Isaka2013Few-Body-Systems54.1219}. The multidimensionally constrained RMF model has been used to study the shape evolution of hypernuclei in the Argon hyperisotopes \cite{Zhou2014PRC89.044307}. The beyond mean-field approach has been utilized to discuss the evolution of $p$-state energies and composition in the Carbon hyperisotopes \cite{Xia2017Sci.China-Phys.Mech.Astron60.102021}, as well as the hyperon halo structures in the Boron and Carbon hyperisotopes \cite{Zhang2021PRC103.034321, Xue2022PRC106.044306}. The studies exhibit the significance of isospin role in the description of hypernuclear structure. In fact, with the development of hypernuclear spectroscopy, new experiments related to hypernuclei have been initiated, such as the planned measurements in the J-PARC project, aiming to study the $ \Lambda $ hyperon binding energies in neutron-rich hyperisotopes of $^{124-136}_\Lambda$Sn \cite{Choi2022EPJA58.161, Aoki2021arXive2110.04462}. These experiments will provide crucial information about the properties of hypernuclei associated with various isospin circumstance.

In view of the essential role of nuclear in-medium effects on hypernuclear structure and their relevance to the isotopic evolution, we aim to further expand the density-dependent RMF model to investigate the structure of single-$\Lambda$ hypernuclei in Oxygen hyperisotopes. First, we will introduce the theoretical framework of the hypernuclear RMF approach in Sec. {\ref{Theoretical Framework}}. Then, the induced $\Lambda$-nucleon ($\Lambda N$) effective interactions will be determined by fitting $\Lambda$ separation energies to the experimental data for DD-LZ1 Lagrangian. To give the results and discussion, the influence of nuclear in-medium effects will be studied in Sec. {\ref{Results and Discussion}}, on the isospin dependence of hypernuclear bulk properties, hyperon spin-orbit splitting and matter/hyperon radius. Finally, a summary will be given in Sec. {\ref{Summary and Outlook}}.

\section{DDRMF approach for spherical single-$\Lambda$ hypernuclei}\label{Theoretical Framework}

To describe single-$\Lambda$ hypernuclei within the meson-exchanged type of the relativistic mean-field theory, the covariant Lagrangian density serves as the foundation, which is
\begin{align}\label{eq:Lagrangian}
 \mathscr{L} = \mathscr{L}_B + \mathscr{L}_{\varphi} + \mathscr{L}_I,
\end{align}
where the terms of free fields read as
\begin{align}\label{eq:LagrangianBMA}
\mathscr{L}_B=&\sum_B\bar{\psi}_B\left(i\gamma^\mu\partial_\mu-M_B\right)\psi_B,\\
\mathscr{L}_{\varphi}=&+\frac{1}{2}\partial^\mu\sigma\partial_\mu\sigma-\frac{1}{2}m_\sigma^2\sigma^2-\frac{1}{4}\Omega^{\mu\nu}\Omega_{\mu\nu}+\frac{1}{2}m_\omega^2\omega^\mu\omega_\mu\notag\\&\quad-\frac{1}{4}\vec{R}^{\mu\nu}\cdot\vec{R}_{\mu\nu}+\frac{1}{2}m_\rho^2\vec{\rho}^{\mu}\cdot\vec{\rho}_\mu-\frac{1}{4}F^{\mu\nu}F_{\mu\nu},
\end{align}
where the index $B$ ($B'$) represents nucleon $N$ or hyperon $\Lambda$, with its sum $\sum_B$ over nucleon $N$ and hyperon $\Lambda$. The masses of the baryon and mesons are given by $M_B$ and $m_\phi$ ($ \phi=\sigma, \omega^{\mu}, \vec{\rho}^{\mu} $), while $\Omega^{\mu\nu}$, $\vec{R}^{\mu\nu}$ and $F^{\mu\nu}$ are the field tensors of vector mesons $\omega^{\mu}, \vec{\rho}^{\mu}$ and photon $A^\mu$, respectively. The interaction between nucleon (hyperon) and mesons (photon) is involved by the Lagrangian $\mathscr{L}_I$,
\begin{align}\label{eq:LagrangianI}
\mathscr{L}_I=\sum_B\bar{\psi}_B&\left(-g_{\sigma B}\sigma-g_{\omega B}\gamma^\mu\omega_\mu\right)\psi_B\nonumber\\
+\bar{\psi}_N&\left(-g_{\rho N}\gamma^\mu\vec{\tau}\cdot\vec{\rho}_\mu-e\gamma^\mu\frac{1-\tau_3}{2}A_\mu\right)\psi_N.
\end{align}
Here the $\Lambda$ hyperon (namely $\psi_B$ taken as $\psi_\Lambda$), which is charge neutral with isospin zero, only takes part in interactions that are spread by isoscalar mesons. The nuclear in-medium effects are introduced phenomenologically via the coupling strengths $g_{\phi B}$ ($g_{\phi N}$), which use baryon-density dependent functions in density-dependent RMF (DDRMF) approach to define the strengths of different meson-baryon (meson-nucleon) couplings \cite{Long2006PLB640.150, Wei2020CPC44.074107}.

The effective Hamiltonian operator for $\Lambda$ hypernuclei can be obtained by performing the general Legendre transformation on the Lagrange density $\mathscr{L}$ in Eq. \eqref{eq:Lagrangian}, and it can be written as the sum of the kinetic energy operator $\hat{T}$ and the potential energy operator $\hat{V}_{\varphi}$,
\begin{align}\label{eq:Hamiltonian}
\hat{H}\equiv&~\hat{T}+\sum_{\varphi}\hat{V}_\varphi\notag\\
=&\int dx \sum_{B} \bar{\psi}_{B}(x)(-i\bm{\gamma}\cdot\bm{\nabla}+M_{B}) \psi_{B}(x)\notag\\
&+ \frac{1}{2} \int dx \sum_{B} \sum_{\varphi} \left[\bar{\psi}_{B} \mathscr{G}_{\varphi B} \psi_{B}\right]_{x} D_{\varphi}(x,x') \left[\bar{\psi}_{B'} \mathscr{G}_{\varphi B'} \psi_{B'}\right]_{x'},
\end{align}
here $x$ is four-vector $(t,\bm{x})$. Correspondingly, we define interaction vertices $\mathscr{G}_{\varphi B}(x)$ for a various of meson (photon)-nucleon (hyperon) coupling channels, which for isoscalar $\sigma$ and $\omega$ mesons are represented as
\begin{subequations}\label{eq:vertice for sigmaomega}
  \begin{align}
    \mathscr{G}_{\sigma B}(x) = &+g_{\sigma B}(x),\\
    \mathscr{G}_{\omega B}^\mu(x) = &+g_{\omega B}(x)\gamma^{\mu}.
  \end{align}
\end{subequations}
Notably, both nucleons and the $\Lambda$ hyperon can contribute to the isoscalar meson fields. However, for the remaining isovector mesons and photon fields, it is expected that their interaction vertices solely connect to nucleons since the isoscalar and charge-zero nature of $\Lambda$ hyperon,
\begin{subequations}\label{eq:vertice for rhopiA}
  \begin{align}
    \mathscr{G}_{\rho N}^\mu(x) = &+g_{\rho N}(x) \gamma^{\mu} \vec{\tau},\\
    \mathscr{G}_{A N}^\mu(x) = &+e\gamma^{\mu}\frac{1-\tau_{3}}{2}.
  \end{align}
\end{subequations}
As the retardation effects could be neglected in the majority of RMF models, the meson (photon) propagators $ D_{\phi} $ ($ D_{A} $) read as
\begin{align}
D_{\phi}(\bm{x},\bm{x}')=\frac{1}{4\pi}\frac{e^{-m_\phi|\bm{x}-\bm{x}'|}}{|\bm{x}-\bm{x}'|},\quad
D_A(\bm{x},\bm{x}')=\frac{1}{4\pi}\frac{1}{|\bm{x}-\bm{x}'|}.
\end{align}

The baryons field operator $\psi_{B}$ in the Hamiltonian \eqref{eq:Hamiltonian} can be second quantized in the positive-energy space under the no-sea approximation as
  \begin{align}
    \psi_B(x)
    &=\sum_if_i(\bm{x})e^{-i\epsilon_i t}c_i.\label{eq:fi}
  \end{align}
Here, $f_i$ represents the Dirac spinor, while $c_i$ denote the annihilation operators for state $i$. Accordingly, the energy functional $E$ is determined by evaluating the expectation value of the Hamiltonian with respect to a trial Hartree-Fock ground state $|\Phi_0\rangle$,
\begin{align}
E & = \left\langle\Phi_{0}|\hat{H}| \Phi_{0}\right\rangle  = \left\langle\Phi_{0}|\hat{T}| \Phi_{0}\right\rangle+\sum_{\varphi}\left\langle\Phi_{0}\left|\hat{V}_{\varphi}\right| \Phi_{0}\right\rangle.
\end{align}
Then the binding energy of a $\Lambda$ hypernucleus is written by
\begin{align}
  E=&\sum_{B}(E_{\rm{kin},B} + E_{\sigma,B} + E_{\omega,B}) + E_{\rho,N}+E_{\rm{e.m.}} + E_{\rm{c.m.}} + E_{\rm{pair}},
  \label{eq:Etot}
\end{align}
where the kinetic energy functional of baryons is shown by $E_{\rm{kin},B}$. The contributions of the potential energy functional from $\sigma$ and $\omega$ are denoted by the variables $E_{\sigma,B}$ and $E_{\omega,B}$. Additionally, $E_{\rho,N}$ and $E_{\rm{e.m.}}$ are used to represent the contributions from $\rho$ and $A$, respectively. The center-of-mass adjustment to the mean-field is represented by the term $E_{\rm{c.m.}}$, while $E_{\rm{pair}}$ takes into account the contribution from nucleon pairing correlations \cite{Ding2022PRC106.054311}.

The role of deformation in single-$\Lambda$ hypernuclei has been discussed in various density functional models \cite{Lu2011PRC84.014328, Xia2017Sci.China-Phys.Mech.Astron60.102021, Xia2023Sci.China-Phys.Mech.Astron66.252011, Xue2023PRC107.044317}, which may generate non-negligible effects on the single-particle energies like in Carbon hyperisotopes \cite{Lu2011PRC84.014328, Xia2017Sci.China-Phys.Mech.Astron60.102021, Xue2023PRC107.044317}. To describe single-$\Lambda$ hypernuclei, in particularly the Oxygen hyperisotopes discussed hereafter, we just restrict the RMF approach to the spherical symmetry. Correspondingly, the Dirac spinor $f_i(\bm{x})$ of the nucleon or hyperon in Eq. \eqref{eq:fi} has the following form:
\begin{align}
  f_{n\kappa m}(\bm{x}) =  \frac{1}{r} \left(\begin{array}{c}iG_a(r)\Omega_{\kappa m}(\vartheta,\varphi)\\ F_a(r)\Omega_{-\kappa m}(\vartheta,\varphi) \end{array}\right),
\end{align}
where the index $a$ consists of the set of quantum numbers $(n\kappa) = (njl)$, and $\Omega_{\kappa m}$ is the spherical spinor. Meanwhile, the propagators can be expanded in terms of spherical Bessel and spherical harmonic functions as
\begin{align}\label{eq:propagator}
D_{\phi}(\bm{x},\bm{x}^{\prime}) = \sum_{L=0}^{\infty}\sum_{M=-L}^{L}(-1)^{M}R^{\phi}_{LL}\left( r, r^{\prime}\right) Y_{LM}\left(\bm{\Omega}\right)Y_{L-M}\left(\bm{\Omega}^{\prime}\right),
\end{align}
where $\bm{\Omega}=(\vartheta,\varphi)$, and $R_{LL}$ contains the modified Bessel functions $I$ and $K$ as
\begin{align}
  R_{L L}^{\phi}\left(r, r^\prime\right)&=\sqrt{\frac{1}{rr^{\prime}}} I_{L+\frac{1}{2}}\left(m_{\phi}r_{<}\right) K_{L+\frac{1}{2}}\left(m_{\phi}r_{>}\right),\\
  R_{L L}^{A}\left(r, r^\prime\right)&=\frac{1}{2L+1}\frac{r_{<}^{L}}{r_{>}^{L+1}}.
\end{align}

In the DDRMF approach, the meson-baryon coupling strengths are adopted as a function of baryon density $\rho_{b}$, which are written by
\begin{align}\label{eq:coupling_constants}
g_{\phi B}\left(\rho_{b}\right)=g_{\phi B}(0) f_{\phi B}(\xi) \quad\text{or}\quad
g_{\phi B}\left(\rho_{b}\right)=g_{\phi B}(0) e^{-a_{\phi B} \xi},
\end{align}
where $\xi=\rho_{b}/\rho_{0}$ with $\rho_{0}$ the saturation density of nuclear matter, and
\begin{align}
  f_{\phi B}(\xi)=a_{\phi B}\frac{1+b_{\phi B}(\xi+d_{\phi B})^2}{1+c_{\phi B}(\xi+d_{\phi B})^2}.
\end{align}
The free coupling strength at $\rho_{b}=0$ is represented by $g_{\phi B}(0)$ in the expression above. To keep the variational self-consistency between the energy density functional and single-particle properties, the extra terms in baryon self-energies, namely the rearrangement terms, will occur due to the density dependence of the coupling strengths. The single-particle (nucleon or hyperon) properties can be determined by solving the Dirac equation,
\begin{align}\label{eq:Dirac}
\varepsilon_{a,B}
\begin{pmatrix}
G_{a,B}(r) \\ F_{a,B}(r)
\end{pmatrix} =&
\begin{pmatrix}
\Sigma_+^B(r) & \displaystyle-\frac{d}{dr}+\frac{\kappa_{a,B}}{r} \\
\displaystyle\frac{d}{dr}+\frac{\kappa_{a,B}}{r} & -\left[2M_{B}-\Sigma_-^B(r)\right]
\end{pmatrix}
\begin{pmatrix}
G_{a,B}(r) \\ F_{a,B}(r)
\end{pmatrix}.
\end{align}
Here the self-energies $\Sigma_\pm^B=\Sigma_{0,B}\pm\Sigma_{S,B}$ composed by the vector and scalar terms. The scalar self-energy $\Sigma_{S,B} = \Sigma_{S,B}^{\sigma}$, and the time component of the vector one has
\begin{align}
  \Sigma_{0,B}(r) = \sum_{\phi}\Sigma_{0,B}^{\phi}(r)+\Sigma_{R}(r),
  \label{eq:Sig0}
\end{align}
where $\phi=\omega, \rho$ for nucleons, and $\phi=\omega$ for $\Lambda$ hyperon. The self-energies of nucleon or hyperon include scalar one $\Sigma_{S,B}$ and vector one $\Sigma_{0,B}$, in which the coupling of isoscalar mesons contributes as follows,
\begin{subequations}
\begin{align}
 \Sigma_{S,B}^{\sigma}(r)&=-g_{\sigma B}(r)\sum_{B^\prime}\int r^{\prime2}dr^\prime g_{\sigma B^\prime}(r^\prime)\rho_{s,B^\prime}(r^\prime)R^{\sigma}_{00}(r,r^\prime),\label{eq:Sigma_S,B}\\
 \Sigma_{0,B}^{\omega}(r)&=+g_{\omega B}(r)\sum_{B^\prime}\int r^{\prime2}dr^\prime g_{\omega B^\prime}(r^\prime)\rho_{b,B^\prime}(r^\prime)R^{\omega}_{00}(r,r^\prime).\label{eq:Sigma_0,B}
\end{align}
\end{subequations}
Here, $\rho_{s,B}$ and $\rho_{b,B}$ represent the scalar and baryon density, respectively \cite{Ding2022PRC106.054311}. Additionally, the rearrangement term $\Sigma_R$ appears in DDRMF approach, which contain the summation over all baryons for the isoscalar case of $\phi=\sigma,\omega$, but only over nucleons for the isovector one. For example, the contribution from $\sigma-S$ coupling is shown as
\begin{align}
\Sigma_{R,\sigma}(r)=\sum_{B}\frac{1}{g_{\sigma B}} \dfrac{\partial g_{\sigma B}}{\partial \rho_{b}}\rho_{s,B}\Sigma_{S,B}^{\sigma}(r).
\label{eq:Erea}
\end{align}

\section{Results and Discussion}\label{Results and Discussion}

In recent years, there has been extensive theoretical research on hypernuclei, particularly focusing on the simplest single-$ \Lambda $ hypernuclei, using RMF and RHF theories. In this section, we aim to extend the effective interaction DD-LZ1 \cite{Wei2020CPC44.074107}, which has been proven to be successful and promising in determining the properties of nuclear structure in both bulk and single-particle aspects, to incorporate $\Lambda$ hyperon within the framework of RMF model. To give a comparative study and illustrate the role of nuclear in-medium effects, the calculations with DD-LZ1 will be accompanied by several existing effective $ \Lambda N $ interactions within CDF models. These interactions have been significantly expanded to incorporate the degrees of freedom of the $\Lambda$ hyperon and have yielded many successful findings in the study of hypernuclear structure and the properties of dense stars. In detail, density-dependent RMF effective interactions DD-LZ1 \cite{Wei2020CPC44.074107}, PKDD \cite{Ding2022PRC106.054311}, DD-ME2, TW99, DDV \cite{Tu2022APJ925.16}, density-dependent RHF (DDRHF) effective interactions PKO1, PKO2, PKO3 \cite{Ding2022PRC106.054311}, and nonlinear RMF (NLRMF) effective interactions NL-SH \cite{Mares1994PRC49.2472} and PK1 \cite{Ren2017PRC95.054318} were selected. In these CDF functionals, the $\omega$-tensor coupling which has been proved to be essential in reducing $\Lambda$'s spin-orbit splitting in hypernuclei \cite{Jennings1990PLB246.1990325} is ignored. The Dirac equation is solved in a radial box size of $R=20$ fm with a step of 0.1 fm. For open-shell hypernuclei, we employ the BCS method to account for pairing correlations. As the strength of hyperon pairing correlations remains uncertain and may become essential in multi-$\Lambda$ hypernuclei, our current work solely considers pairing correlations between $ nn $ and $ pp $ pairs by using the finite-range Gogny force D1S \cite{Berger1984NPA428.23}, see Refs. \cite{Meng1998NPA.635.3, Long2010PRC81.024308, Geng2020PRC101.064302, Geng2022PRC105.034329} for details. In addition, the blocking effect should be taken into account for the last valence nucleon or hyperon, with a detailed description to the Ref. \cite{Ding2022PRC106.054311}.

\begin{figure}[hbpt]
 \centering
 \includegraphics[width=0.48\linewidth]{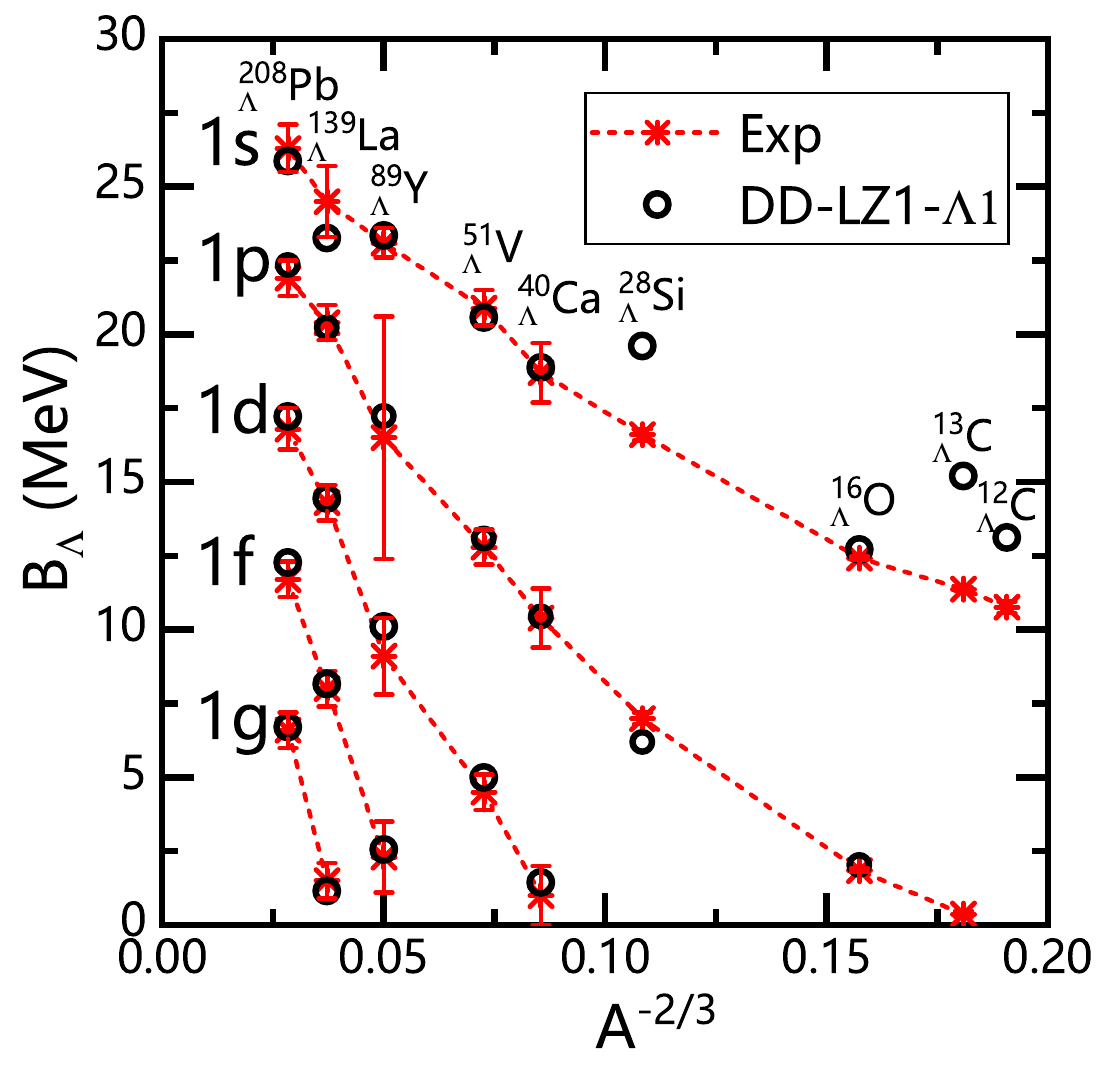}
 \caption{(Color Online) The calculated $\Lambda$ separation energies $B_\Lambda$ for the single-$\Lambda$ hypernuclei with the RMF effective interaction DD-LZ1-$\Lambda1$ in comparison with the experimental data taken from Ref. \cite{Pile1991PRL66.2585, Gal2016Rev.Mod.Phys.88.035004}.}\label{Fig:BL_DDLZ1}
\end{figure}

\subsection{Density dependence of $\Lambda N$ effective interaction}\label{separation energies}

For the theoretical study of hypernuclear structure, the $ \Lambda N $ interaction must be determined first. Since the $ \Lambda $ hyperon is an electrically neutral particle with isospin zero, our focus lies on the coupling strengths between the isoscalar-scalar $ \sigma $ meson and the isoscalar-vector $ \omega $ meson with the $ \Lambda $ hyperon. For convenience, we introduce the ratio of the coupling strengths between the meson-hyperon and meson-nucleon, $g_{\phi\Lambda}/g_{\phi N}$. According to the n\"{a}ive quark model \cite{Dover1984PPNP12.171}, we fix the ratio of the isoscalar-vector meson coupling strength $g_{\omega\Lambda}/g_{\omega N}$ to 0.666, while the ratio of the isoscalar-scalar one $ g_{\sigma\Lambda}/g_{\sigma N} $ can be obtained by reproducing the $ \Lambda $ hyperon separation energy $ B_{\Lambda} $ experimental data for $^{16}_\Lambda$O, $^{40}_\Lambda$Ca, and $^{208}_\Lambda$Pb \cite{Pile1991PRL66.2585, Gal2016Rev.Mod.Phys.88.035004}. In the fitting process, the hyperon is placed in the $ 1s_{1/2} $ ground state, and the $ B_{\Lambda} $ is defined as follows:
\begin{align}
    B_\Lambda(^{A}_\Lambda Z) = E(^{A-1}Z) - E(^{A}_\Lambda Z),\label{eq:BLmd}
\end{align}

Based on the effective interaction DD-LZ1, we finally obtained a new set of $\Lambda N$ interaction, namely DD-LZ1-$\Lambda1$, after a fitting process of Levenberg-Marquardt minimization. Then, we calculated the $ \Lambda $ separation energy $ B_{\Lambda} $ as well as the single-$\Lambda$ energy, with hyperon occupying the ground state $1s_{1/2}$ or possible excited states with higher angular momentum $l_\Lambda$. For $ B_{\Lambda} $ of DD-LZ1-$ \Lambda1 $, a remarkable agreement with experimental data is found for most of hypernuclei, except for $^{28}_\Lambda$Si with significant deformation and Carbon hyperisotopes with light mass, as shown in Fig. \ref{Fig:BL_DDLZ1}. Actually, more accuracy description to the light-mass Carbon hyperisotopes could be obtained, by limiting the mass region of fitting and taking into account the deformation effects \cite{Rong2021PRC104.054321}. To investigate the deviation in describing the structural properties of single-$\Lambda$ hypernuclei using different CDF effective interactions, the coupling strength of DD-LZ1-$\Lambda1$ in comparison with other selected CDF functionals are listed in Table \ref{Tab:Coupling_Constants}. One could check the root-mean-square deviation $\Delta$ for $ B_{\Lambda} $ between theoretical calculation and experimental value, which is defined by
\begin{equation}
\Delta\equiv\sqrt{\frac{1}{N}\sum\limits_{i=1}^N(B_{\Lambda,i}^{\rm{exp.}}-B_{\Lambda,i}^{\rm{cal.}})^2}.
\end{equation}
To reveal the systematics, we define $\Delta_{1}$ to be the deviation only for $^{16}_\Lambda$O, $^{40}_\Lambda$Ca, and $^{208}_\Lambda$Pb, as well as $\Delta_{2}$ that suitable for all hypernuclei.

\begin{table}[t]
  \centering
  \caption{The $\sigma$-$\Lambda$ coupling strengths $g_{\sigma\Lambda}/g_{\sigma N}$ fitted for the DDRMF effective interactions DD-LZ1-$ \Lambda1 $, PKDD-$ \Lambda1 $ \cite{Ding2022PRC106.054311}, DD-ME2, TW99 and DDV \cite{Tu2022APJ925.16}, the DDRHF ones PKO1-$ \Lambda1 $, PKO2-$ \Lambda1 $ and PKO3-$ \Lambda1 $ \cite{Ding2022PRC106.054311}, as well as NLRMF ones NL-SH \cite{Mares1994PRC49.2472} and PK1 \cite{Ren2017PRC95.054318} by minimizing the root-mean-square deviation $\Delta_{1}$ (in MeV) from the experiment values of $\Lambda$ separation energies of $^{16}_\Lambda$O, $^{40}_\Lambda$Ca and $^{208}_\Lambda$Pb, where the $\omega$-$\Lambda$ coupling is fixed to be $g_{\omega\Lambda}/g_{\omega N} = 0.666$. $ \Delta_{2} $ represent the root-mean-square deviation between the theoretical calculations and experimental values of $\Lambda$ separation energies for all hypernuclei shown in Fig. \ref{Fig:BL_DDLZ1}.
  }\label{Tab:Coupling_Constants}
  \renewcommand{\arraystretch}{1.5}
  \setlength{\tabcolsep}{4.5pt}
  \begin{tabular*}{\textwidth}{ccccccccccc}
\hline\hline
                                 & DD-LZ1-$\Lambda$1 & PKDD-$\Lambda$1 & DD-ME2 & TW99  & DDV   & PKO1-$\Lambda$1 & PKO2-$\Lambda$1 & PKO3-$\Lambda$1 & NL-SH & PK1    \\ \hline
$g_{\sigma\Lambda}/g_{\sigma N}$ & 0.615             & 0.620           & 0.620  & 0.617 & 0.622 & 0.596           & 0.591           & 0.594           & 0.621 & 0.618  \\
$\Delta_{1}$                     & 0.319             & 0.363           & 0.245  & 0.375 & 0.473 & 0.265           & 0.260           & 0.407           & 0.916 & 0.519  \\
$\Delta_{2}$                     & 1.810             & 0.734           & 0.710  & 0.684 & 3.460 & 0.683           & 0.527           & 0.881           & 1.614 & 1.184  \\ \hline
\hline
\end{tabular*}
\end{table}

From Table \ref{Tab:Coupling_Constants}, it can be seen that different CDF theoretical models have good descriptions for $^{16}_\Lambda$O, $^{40}_\Lambda$Ca and $^{208}_\Lambda$Pb, and most parameter sets have good consistency for hypernuclear theoretical calculations and experimental data over a large mass range from $^{12}_\Lambda$C to $^{208}_\Lambda$Pb. In addition, by comparing three different types of CDF effective interactions, we can find that when the ratio of isospin scalar-vector meson coupling strength is fixed to the same value, the ratio of isospin scalar-scalar meson coupling strength $ g_{\sigma\Lambda}/g_{\sigma N} $ may satisfy certain linear correlations with the ratio of isospin scalar-vector meson coupling strength, which has been systematically explored in some works \cite{Wang2013Com.Theor.Phys.60.479,Rong2021PRC104.054321,Sun2023APJ942.55}. It should be pointed out that the linear correlation of meson-hyperon coupling strength ratios obtained in the RMF framework is obviously not suitable for density-dependent RHF models \cite{Ding2022PRC106.054311}.

In DDRMF approach, the in-medium effects of nuclear force are effectively embedded in the density-dependent shape of meason-baryon coupling strength, playing the role in the nuclear structure via the equilibrium of nuclear dynamics from various coupling channels. In recent years, analysis based on the equilibrium of nuclear in-medium dynamics has been applied to clarify the mechanism of the pseudospin symmetry, the shell evolution, the liquid-gas phase transition, and hyperon's spin-orbit splitting in the CDF models \cite{Liu2020PLB806.135524, Geng2020PRC101.064302, Wei2020CPC44.074107, Yang2021PRC103.014304, Ding2022PRC106.054311}. The delicate in-medium balance between nuclear attractive and repulsive interactions may be significantly altered by treating the density dependence of coupling strength differently, impacting the description of the properties of nuclear matter and finite nuclei with different CDF effective interactions.
\begin{figure}[htbp]
  \centering
  \includegraphics[width=0.48\linewidth]{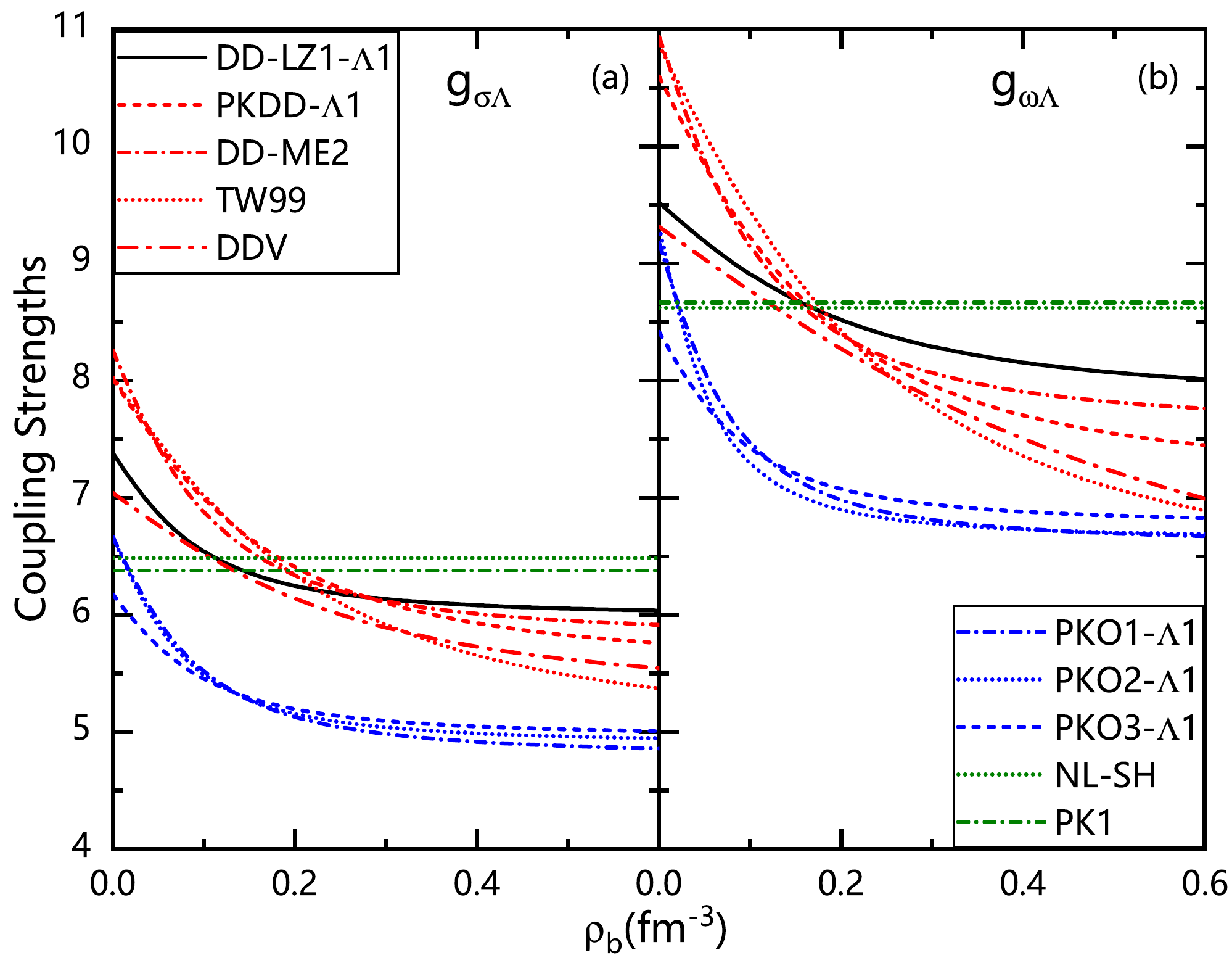}
  \caption{(Color Online) Meson-hyperon coupling strengths, namely, the isoscalar $ g_{\sigma\Lambda} $ [panel (a)]  and $ g_{\omega\Lambda} $ [panel (b)], as functions of baryonic density $ \rho_{b} $(fm$^{-3}$)  for the DDRMF effective interactions DD-LZ1-$ \Lambda1 $, PKDD-$ \Lambda1 $, DD-ME2, TW99 and DDV, the DDRHF ones PKO1-$ \Lambda1 $, PKO2-$ \Lambda1 $ and PKO3-$ \Lambda1 $, as well as NLRMF ones NL-SH and PK1.}\label{Fig:Coupling_constant}
\end{figure}

To provide a comprehensive understanding of the in-medium equilibrium in hypernuclei, we present the density dependence of coupling strengths for selected CDF effective interactions in Fig. \ref{Fig:Coupling_constant}(a) and Fig. \ref{Fig:Coupling_constant}(b), corresponding to the isoscalar-scalar channel $ g_{\sigma\Lambda} $ and isoscalar-vector one $ g_{\omega\Lambda} $. There are systematic divergences of the meson-hyperon coupling strengths with increasing density among density-dependent RMF, density-dependent RHF, and nonlinear RMF effective interactions. Notably, the density dependence of $g_{\sigma\Lambda}$ and $g_{\omega\Lambda}$ is significantly reduced in the DDRHF effective interaction compared to the DDRMF effective interaction. This pronounced reduction in density dependence also influences the description of single-particle properties in hypernuclei, such as $ \Lambda $ hyperon spin-orbit splitting \cite{Ding2022PRC106.054311}. Furthermore, in contrast to density-dependent interactions, the NLRMF effective interaction exhibits density-independent characteristics for $g_{\sigma\Lambda}$ and $g_{\omega\Lambda}$. Consequently, when applying these three types of CDF effective interactions to single-$\Lambda$ hypernuclei, the systematic deviation could take place in describing the isospin dependence of the hypernuclear structure.

\subsection{Bulk properties of single-$\Lambda$ hypernuclei in Oxygen hyperisotopes}
To focus on the isospin dependence of single-particle properties, we choose the $\Lambda$ hypernuclei and their nucleonic counterpart in Oxygen (hyper)isotopes as examples, since they usually take the spherical symmetry. To check the accuracy of the chosen interactions in describing the properties of finite nuclei, we first calculated the binding energies $E_{B}$, charge radii $R_{c}$, and matter radii $R_{\rm{m}}$ for Oxygen isotopes using the DD-LZ1 effective interaction. We compared the theoretical calculations with experimental measurements, which were taken from Refs. \cite{Wang2021CPC45.030003,Zhang2022ADNDT144.101488,Kaur2022PRL129.142502}. From the results in Table \ref{Tab:Nuclei}, we can see that the theoretical calculations and experimental measurements are in good agreement for both the binding energies $E_{B}$ and the charge radii $R_{c}$, for the interaction DD-LZ1. It is worth noting that the total matter radius $R_{\rm{m}}$ of finite nuclei, unlike the charge radius, still has significant uncertainties based on heavy ion reaction experiments. The theoretical calculations of $R_{\rm{m}}$ reconcile with the experimental measurements with the existence of error bars.

\begin{table}[hbpt]
  \centering
  \caption{Binding energy $ E_{B} $, charge radii $ R_{\rm{c}} $ and matter radii $ R_{\rm{m}} $ of normal nuclei $^{Z+N}$O, calculated by DDRMF effective interaction DD-LZ1, compared to the experimental data \cite{Angeli2013ADNDT99.69, Li2021ADNDT140.101440, Wang2021CPC45.030003, Kaur2022PRL129.142502}.}
  \label{Tab:Nuclei}
  \setlength{\tabcolsep}{15pt}
  \renewcommand{\arraystretch}{1.25}
\begin{tabular}{ccccccc}
\hline\hline
  Nucleus &
  $E_{B}$(MeV) &
  $E_{B}^{\rm{exp.}}$(MeV) &
  $R_{\rm{c}}$(fm) &
  $R_{\rm{c}}^{\rm{exp.}}$(fm) &
  $R_{\rm{m}}$(fm) &
  $R_{\rm{m}}^{\rm{exp.}}$(fm)   \\ \hline
  $^{14}$O    &  -99.699  &  -98.732  & 2.766  &         & 2.543  &             \\
  $^{16}$O    & -128.215  & -127.619  & 2.752  & 2.699   & 2.619  & 2.57(2)     \\
  $^{18}$O    & -140.017  & -139.808  & 2.749  & 2.773   & 2.761  & 2.64(8)     \\
  $^{20}$O    & -150.687  & -151.371  & 2.746  &         & 2.868  & 2.71(3)     \\
  $^{22}$O    & -160.364  & -162.028  & 2.746  &         & 2.955  & 2.90(5)     \\
  $^{24}$O    & -168.802  & -168.960  & 2.761  &         & 3.054  & 3.18(12)    \\
 \hline\hline
\end{tabular}
\end{table}

\begin{table}[t]
  \centering
  \caption{Properties of single-$ \Lambda $ states in hypernucleus $^{Z+N+\Lambda}_{\Lambda}$O calculated with the DDRMF effective interaction DD-LZ1-$ \Lambda1 $, including single-particle energies $ \varepsilon_{\rm{s.p.}} $, binding energies $ E_{B} $, charge radii $ R_{c} $, hyperon radii $ R_{\Lambda} $, hypernuclear matter radii $R_{\rm{m}}$ and $ B(E1) $ value of the transition from the excited $ (\Lambda1p) $ state to the ground $ (\Lambda1s) $ state.}\label{Tab:Hypernuclei_O}
  \setlength{\tabcolsep}{10pt}
  \renewcommand{\arraystretch}{1.25}
\begin{tabular}{cccccccc}
\hline\hline
  Nucleus &
  $\Lambda(nlj)$ &
  $\varepsilon_{\rm{s.p.}}$(MeV) &
  $E_{B}$(MeV) &
  $R_{\rm{c}}$(fm) &
  $R_{\Lambda}$(fm) &
  $R_{\rm{m}}$(fm) &
  $B(E1)$($e^{2}\rm{fm}^2$) \\ \hline
  \multirow{3}{*}{$^{15}_{\Lambda}$O}   & \textbf{$1s_{1/2}$}  & -14.330  & -113.245  &  2.716   & 2.115  & 2.458   &        \\
                                        & \textbf{$1p_{1/2}$}  & -0.413   &  -99.590  &  2.772   & 5.265  & 2.813   &  0.095 \\
                                        & \textbf{$1p_{3/2}$}  & -1.582   & -101.002  &  2.760   & 4.134  & 2.674   &  0.119 \\
  \multirow{3}{*}{$^{17}_{\Lambda}$O}   & \textbf{$1s_{1/2}$}  & -13.086  & -140.507  &  2.704   & 2.323  & 2.555   &        \\
                                        & \textbf{$1p_{1/2}$}  & -1.059   & -128.927  &  2.756   & 4.609  & 2.780   &  0.109 \\
                                        & \textbf{$1p_{3/2}$}  & -2.278   & -130.307  &  2.746   & 3.963  & 2.711   &  0.121 \\
  \multirow{3}{*}{$^{19}_{\Lambda}$O}   & \textbf{$1s_{1/2}$}  & -14.170  & -153.506  &  2.699   & 2.310  & 2.682   &        \\
                                        & \textbf{$1p_{1/2}$}  & -1.720   & -141.540  &  2.751   & 4.291  & 2.861   &  0.090 \\
                                        & \textbf{$1p_{3/2}$}  & -3.036   & -143.003  &  2.740   & 3.824  & 2.815   &  0.097 \\
  \multirow{3}{*}{$^{21}_{\Lambda}$O}   & \textbf{$1s_{1/2}$}  & -15.394  & -165.477  &  2.695   & 2.295  & 2.773   &        \\
                                        & \textbf{$1p_{1/2}$}  & -2.463   & -153.079  &  2.744   & 4.062  & 2.927   &  0.075 \\
                                        & \textbf{$1p_{3/2}$}  & -3.890   & -154.635  &  2.733   & 3.699  & 2.890   &  0.079 \\
  \multirow{3}{*}{$^{23}_{\Lambda}$O}   & \textbf{$1s_{1/2}$}  & -16.804  & -176.670  &  2.688   & 2.277  & 2.829   &        \\
                                        & \textbf{$1p_{1/2}$}  & -3.285   & -163.703  &  2.737   & 3.882  & 2.977   &  0.063 \\
                                        & \textbf{$1p_{3/2}$}  & -4.841   & -165.374  &  2.725   & 3.582  & 2.943   &  0.066 \\
  \multirow{3}{*}{$^{25}_{\Lambda}$O}   & \textbf{$1s_{1/2}$}  & -17.634  & -185.728  &  2.723   & 2.256  & 2.969   &        \\
                                        & \textbf{$1p_{1/2}$}  & -3.925   & -172.669  &  2.757   & 3.836  & 3.079   &  0.052 \\
                                        & \textbf{$1p_{3/2}$}  & -5.522   & -174.326  &  2.748   & 3.562  & 3.055   &  0.055 \\
 \hline\hline
\end{tabular}
\end{table}

Furthermore, we summarize in Table \ref{Tab:Hypernuclei_O} the systematics of the occupied energy level of $ \Lambda $ hyperon, the single-particle energies of $ \Lambda $ hyperon, the total binding energies, the charge radii, and the matter radii of hypernuclei in Oxygen hyperisotopes. In order to give possible reference to hypernuclear experiments, we also calculated the strength of electric dipole transition $ B(E1) $ between the $ \Lambda1p $ and $ \Lambda1s $ occupation states. The transition strength is expressed as
\begin{align}\label{eq:BE1}
B(E1;J_{i}\longrightarrow J_{f})=\frac{3e^{2}_{\Lambda}}{4\pi}\langle f|r|i\rangle^{2}(2j_{f}+1)
\begin{pmatrix}
j_{f} & 1 & j_{i} \\
-\frac{1}{2} & 0 & \frac{1}{2}
\end{pmatrix}^{2},
\end{align}
Where $e_{\Lambda}$ represents the effective charge of the $\Lambda$ hyperon. The integration $\langle f|r|i\rangle$ can be computed using the radial wave functions of the initial and final single-$\Lambda$ state, see Ref. \cite{Zhang2021PRC103.034321} for details.

In the framework of relativistic models, Dirac spinors with both upper and lower components could contribute to determining the value of $ B(E1) $. However, it is checked that the contribution from the lower component is negligible, especially for non-charge exchange channel. Therefore, only the contribution from the upper component is preserved in current calculations as a simplification. The inclusion of $\Lambda$ hyperon causes the so-called impurity effect inside hypernuclei \cite{Hashimoto2006PPNP57.564}. When the $ \Lambda $ hyperon is filled in the $ 1s_{1/2} $ state, we can see from the comparison of the total matter radii in Table \ref{Tab:Hypernuclei_O} and Table \ref{Tab:Nuclei} that the introduction of hyperon causes a shrinkage effect on the hypernuclei, which is approximately $0.06-0.13$ fm. Compared with the ground-state results, we observe a significant enhancement in $ \Lambda $ root-mean-square radii when hyperon is filled in higher-lying $ 1p $ state. This change in the density distribution of hyperon due to different level occupations leads to an overall expansion of the hypernuclear matter radii, different from the $ \Lambda1s $ case. Additionally, with the increase of neutron filling, both the hyperon radii, matter radii and $ B(E_{1}) $ show significant isospin dependence, which can be qualitatively explained by the density-dependence of the coupling strength. As indicated in Table \ref{Tab:Hypernuclei_O}, when $ \Lambda $ hyperon occupies the $ 1p $ state, its density distribution spreads more outward than the nucleonic core. As isospin evolves, more neutrons are filled and their attraction to the hyperon increases, correspondingly leading to a significant reduction in the hyperon radius. For $ B(E_{1}) $, its value is determined not only by the overlap between initial and final states which are sensitive to the neutron number, but also by the effective charge. As a result, the $ B(E_{1}) $ values enlarge a little from $^{15}_{\Lambda}$O to $^{17}_{\Lambda}$O and go down gradually as isospin evolves after $N=8$.

\subsection{Isospin dependence of $\Lambda$ spin-orbit splitting}

\begin{figure}[htbp]
  \centering
  \includegraphics[width=0.75\linewidth]{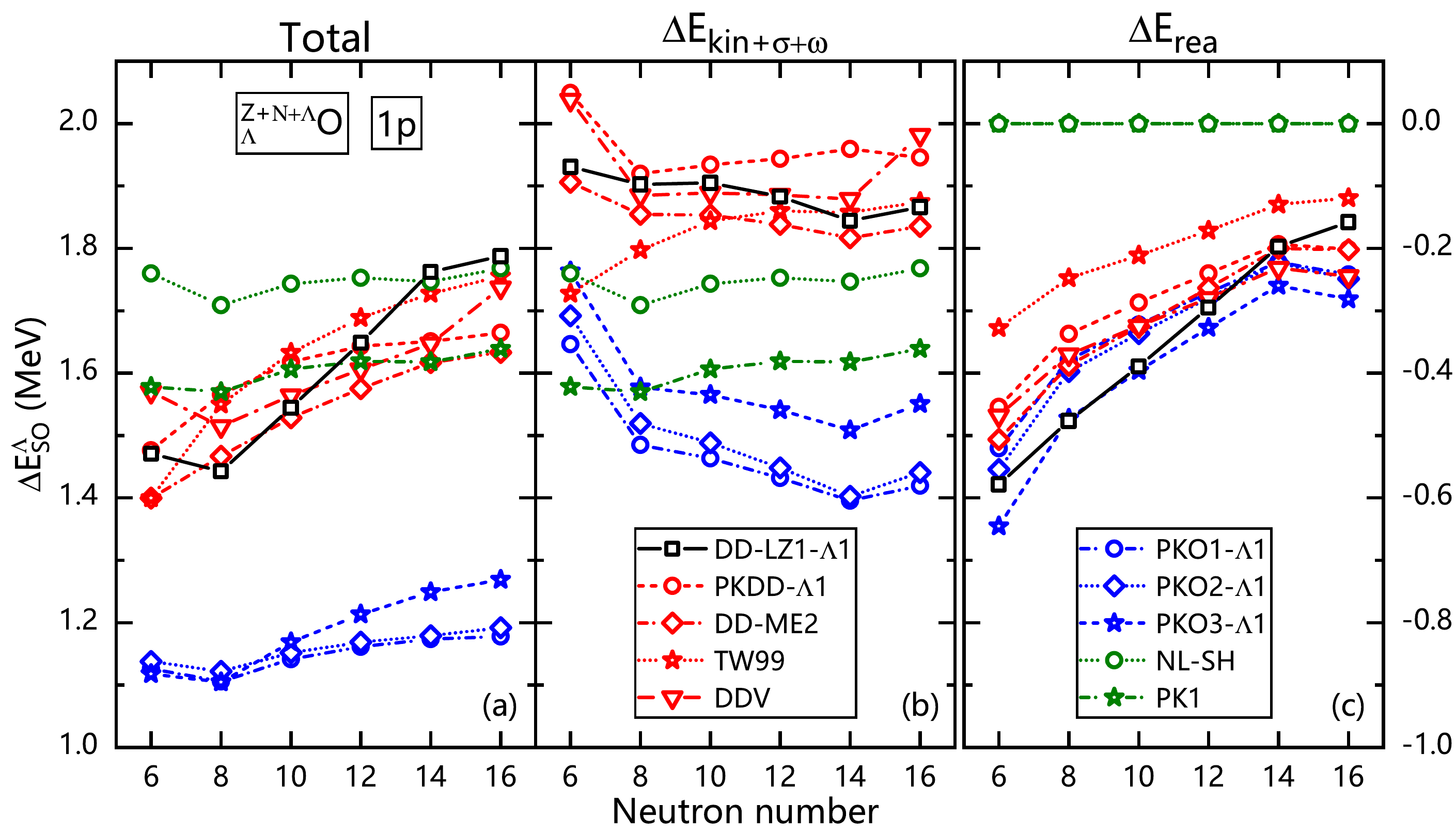}
  \caption{(Color Online) The spin-orbit splitting of $ \Lambda 1p$ spin-partner states as a function of neutron number $N$ for the ground state in $^{Z+N+\Lambda}_{\Lambda}$O hypernuclei [panel (a)], and its contribution $ \Delta E_{\rm{kin}+\sigma+\omega} $ from the sum of the kinetic energy, the density-independent potential energies of $ \sigma $ and $ \omega $ channels [panel (b)], as well as the rearrangement terms $ \Delta E_{\rm{rea}} $ due to density-dependent meson-hyperon couplings [panel (c)]. The results are extracted from the calculations with the DDRMF effective interactions DD-LZ1-$ \Lambda1 $, PKDD-$ \Lambda1 $, DD-ME2, TW99 and DDV, the DDRHF ones PKO1-$ \Lambda1 $, PKO2-$ \Lambda1 $ and PKO3-$ \Lambda1 $, as well as the NLRMF ones NL-SH and PK1.}\label{Fig:Eso_O}
\end{figure}

Motivated by the connection between the density-dependent effective interactions of theoretical models and the isospin-dependent properties of nuclear structure, the spin-orbit splitting of $\Lambda$ hyperon in hypernuclei, as a promising observable in current hypernuclear spectroscopy, will be discussed in this subsection with newly developed DD-LZ1-$\Lambda1$ and other selected CDF functionals. The $\Lambda$'s spin-orbit splitting is defined by the difference of $\Lambda$ single-particle energies between a couple of spin partner states, which is
\begin{align}
\Delta E_{\rm{SO}}^\Lambda \equiv \varepsilon_{j_\Lambda=l_\Lambda-1/2} - \varepsilon_{j_\Lambda=l_\Lambda+1/2}.
\end{align}
As shown in Fig. \ref{Fig:Eso_O}, the analysis is carried out for $ \Lambda $ spin partner states $ 1p $ in Oxygen hyperisotopes, with the $\Lambda$ hyperon occupying its ground state.

In Fig. \ref{Fig:Eso_O}(a), it is seen that the isospin dependence of $\Delta E_{\rm{SO}}^\Lambda$ is clearly distinguished with the chosen CDF functionals. The curves from NLRMF models tend to be stable with increasing neutron number, while for density-dependent RMF or RHF functionals the splitting enlarges generally with isospin. Among them, DD-LZ1-$\Lambda1$ exhibits the most significant isospin dependence. Besides, it is clear that the smaller $ \Lambda $ spin-orbit splitting is predicted by DDRHF compared to RMF, which has been illustrated as a result in single-particle properties since the dynamical equilibrium between nuclear attraction and repulsion is dramatically changed with the appearance of Fock terms \cite{Ding2022PRC106.054311}.

To better understand the evolution of $\Lambda$ spin-orbit splitting with isospin, we could decompose $\Delta E_{\rm{SO}}^\Lambda$ into various parts according to its source of the kinetic or potential energy. The values are obtained by left-multiplying the transferred Dirac spinor to the Dirac equation Eq. \eqref{eq:Dirac}, and separate the integrated contributions from different self-energie terms. For instance, $\Delta E_{\text{rea}}$ comes from the contribution of the rearrangement term $\Sigma_R$ to $\Lambda$ self-energy $\Sigma_{0,\Lambda}$, as seen in Eq. \eqref{eq:Sig0}, due to the density dependence of meson-hyperon couplings. Consequently, the rest one from the kinetic energy and the density-independent potential energies could be summed over, which means $\Delta E_{\text{kin}+\sigma+\omega}\equiv\Delta E_{\rm{SO}}^\Lambda-\Delta E_{\text{rea}}$, as discussed in Fig. \ref{Fig:Eso_O}(b).

It is observed that the values of $\Lambda$ spin-orbit splitting are primarily determined by $\Delta E_{\text{kin}+\sigma+\omega}$. However, the isospin dependence of the splitting is weakly controlled by $\Delta E_{\text{kin}+\sigma+\omega}$ except for $^{15}_\Lambda$O. Attributed to the occupation of $\nu 1p_{1/2}$ orbit, the $\Lambda$ spin-orbit splitting predicted by various CDF functionals systematically reduces from $^{15}_\Lambda$O to $^{17}_\Lambda$O. As has been illustrated in Ref. \cite{Ding2022PRC106.054311}, the spin-orbit coupling potential of hyperon is determined mainly by the radial derivative of the self-energy $\Sigma_-^\Lambda$. In general, the more neutrons are filled into hypernuclei, the larger the density circumstance where the $\Lambda$ hyperon is housing. Thus, if the model is density dependent like DDRMFs and DDRHFs given in Fig. \ref{Fig:Coupling_constant}, the meson-hyperon coupling strength then weakens and $\Delta E_{\rm{SO}}^\Lambda$ should become smaller correspondingly as the neutron number increases. As seen in Fig. \ref{Fig:Eso_O}(b), such a reduction in $\Delta E_{\text{kin}+\sigma+\omega}$ is remarkable from $^{15}_\Lambda$O to $^{17}_\Lambda$O, and relatively less significant at larger neutron numbers.

Different from the NLRMF case, the density-dependent CDFs introduce extra contribution to reinforce the isospin dependence of the splitting, as demonstrated in Fig. \ref{Fig:Eso_O}(c), which cancels the reduction trend in $\Delta E_{\text{kin}+\sigma+\omega}$ overwhelmingly and finally leads to the enhancement of $\Delta E_{\rm{SO}}^\Lambda$ with increasing neutron number in Fig. \ref{Fig:Eso_O}(a). In fact, the contribution $\Delta E_{\text{rea}}$ to $\Lambda$ spin-orbit splitting is originated from the rearrangement terms of $\Lambda$ self-energies $\Sigma_{0,\Lambda}$ which according to Eq. \eqref{eq:Erea} depends on the density slope of the meson-hyperon coupling strength. As the neutron number increases, the density scenario where $\Lambda$ lives could get more intense, consequently weaker density dependence of the meson-hyperon coupling strength, smaller density slope as well as the suppressed value of $\Delta E_{\text{rea}}$. Therefore, the link between the isospin evolution of $\Lambda$ spin-orbit splitting and the in-medium behavior of $\Lambda N$ interaction with baryon density is elucidated from the discussion on Oxygen hyperisotopes. In consequence, possible experimental constraints on $\Delta E_{\rm{SO}}^\Lambda$ along the hyperisotopes could assist us further in understanding the in-medium effects of nuclear force.

\subsection{Isospin dependence of matter and hyperon radii}

\begin{figure}[h]
  \centering
  \includegraphics[width=0.48\linewidth]{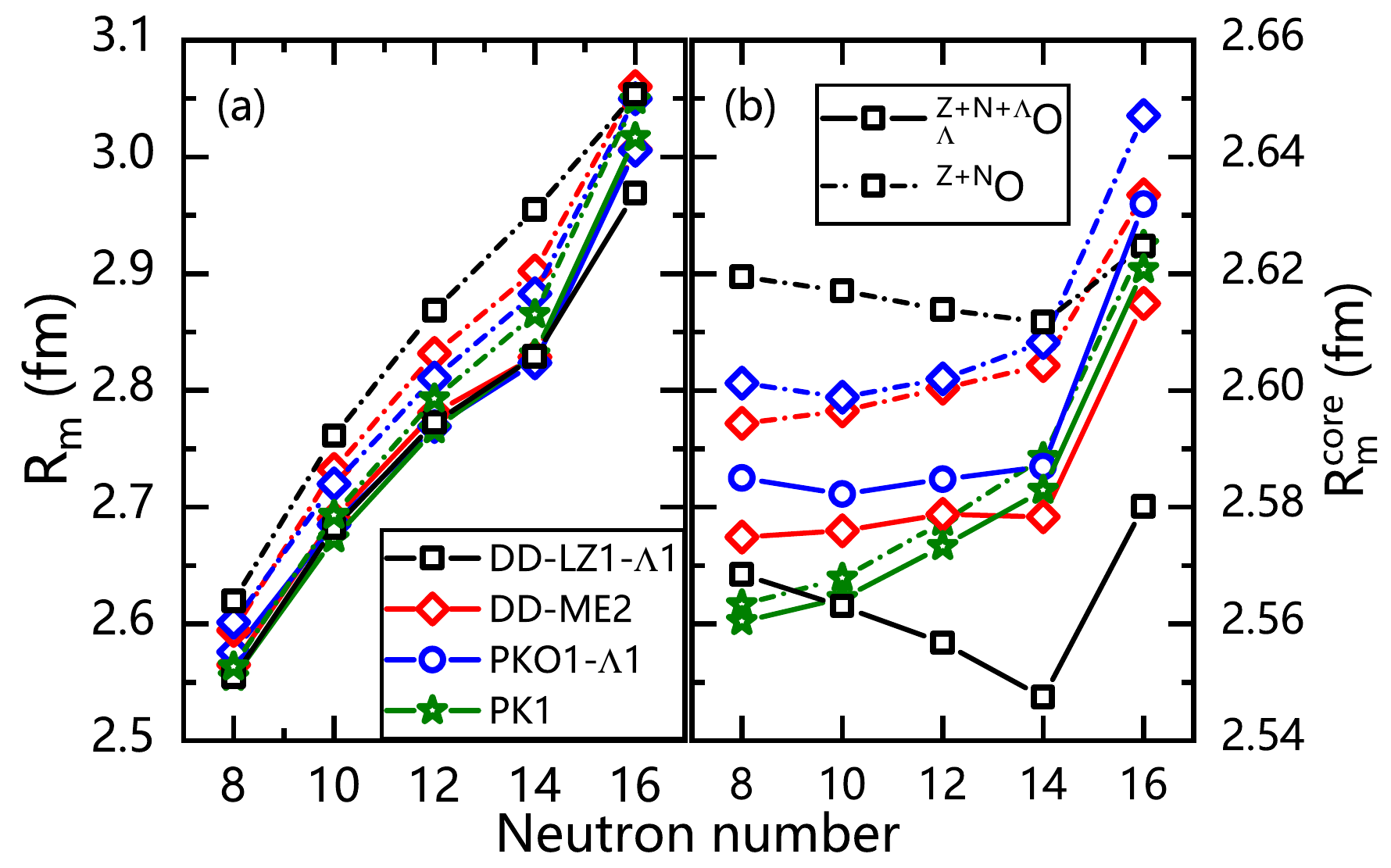}
  \caption{(Color Online) The variation of the matter radii of hypernuclei [Panel (a)] and their $^{16}$O core [Panel (b)] in Oxygen (hyper)isotopes with respect to the neutron number, with the $\Lambda$ hyperon occupying the $1s_{1/2}$ ground state. The solid and dash-dotted lines represent the calculated results for hypernuclei and normal isotopes without hyperon, respectively. The results were obtained with the CDF functionals DD-LZ1-$\Lambda1$, DD-ME2, PKO1-$\Lambda1$ and PK1.}\label{Fig:Rms_tot}
\end{figure}

In the properties of hypernuclear structure, not only the $ \Lambda $ spin-orbit splitting but also the $ \Lambda $ impurity effect could exhibit the information of in-medium nuclear interactions. In Fig. \ref{Fig:Rms_tot}(a), we selected DDRMF functionals DD-LZ1-$\Lambda1$ and DD-ME2, DDRHF's PKO1-$\Lambda1$ and NLRMF's PK1, to illustrate its influence on the matter radii of Oxygen (hyper)isotopes, where the solid and dash-dotted lines correspond to the calculated results for single-$ \Lambda $ hypernuclei and their nucleonic counterpart in Oxygen (hyper)isotopes, respectively. The matter radius $R_m$ in hypernuclei goes up monotonically as the neutron number increases, regardless of the specific model used, where a steep leap from $^{23}_\Lambda$O to $^{25}_\Lambda$O corresponds to the effect of new occupation in $\nu 2s_{1/2}$.

Although divergent values given for Oxygen isotopes without hyperon, all of the selected models are getting closer in size of matter radii for hypernuclei, implying $R_m$ of hypernuclei as a possible model-independent observable. It is evident that the matter radii of Oxygen hyperisotopes contract as compared to their nucleonic counterparts, namely the size shrinkage due to the impurity effect of the $ \Lambda $ hyperon. However, the shrinkage magnitude appears to be strongly model dependent. Among them, the DDRMF effective Lagrangian DD-LZ1-$\Lambda$1 yields the largest difference between the solid and dash-dotted lines, whereas the NLRMF one PK1 shows the smallest disparity. By checking the bulk properties of nuclear matter within these CDFs, it is verified that the shrinkage magnitude correlates well with the incompressibility, which is 230.7 MeV for DD-LZ1, 250.8 MeV for DD-ME2, 250.2 MeV for PKO1, 282.7 MeV for PK1, respectively \cite{Sun2008PRC78.065805, Long2012PRC85.025806, Wei2020CPC44.074107}. In fact, the larger the incompressibility $K$ is, the harder the nucleus is contracted by the exerted attraction from the filled hyperon inside, consequently the weaker size shrinkage effect in the calculated matter radii. The similar relation could be found from the Table II of a work on the isoscalar giant monopole resonance of hypernuclei, where the effective nuclear incompressibility modulus was extracted \cite{Lv2018CPL35.062102}.

\begin{figure}[htbp]
  \centering
  \includegraphics[width=0.48\linewidth]{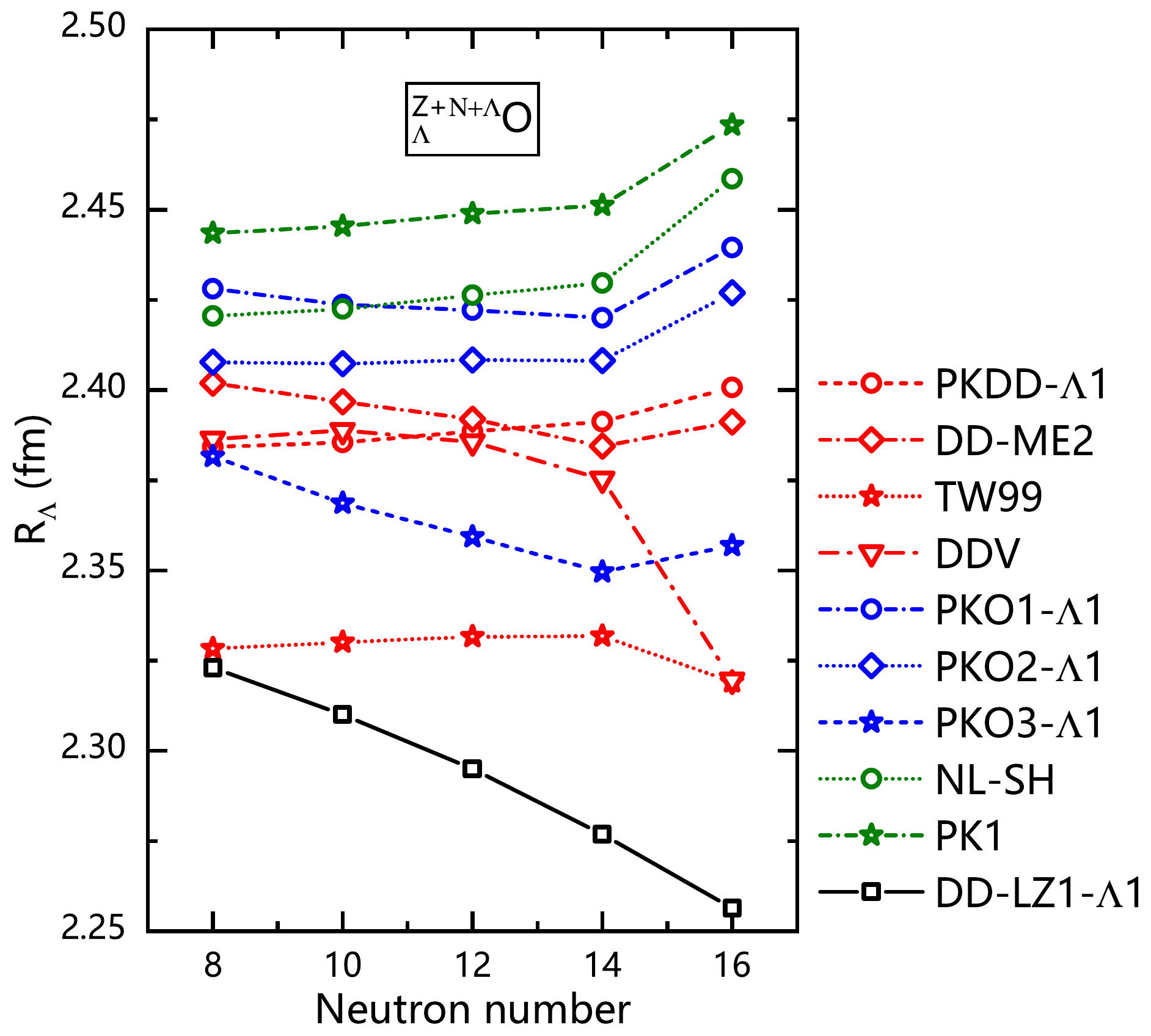}
  \caption{(Color Online) The variation of the $ \Lambda $ hyperon radius with respect to neutron number in Oxygen hyperisotopes is shown. The $ \Lambda $ hyperon is filled in the $1s_{1/2}$ ground state. These results were obtained from calculations using all selected CDF effective interactions.}\label{Fig:Rms_Lambda}
\end{figure}

To further distinguish the effects of different interactions on the description of hypernuclear structure, we investigate the isospin evolution of the $\Lambda$ hyperon radius $R_\Lambda$ in Oxygen hyperisotopes using all selected CDF effective interactions, as shown in Fig. \ref{Fig:Rms_Lambda}. It is seen tangibly that $R_\Lambda$ evolve diversely along Oxygen hyperisotopes with different CDF effective interactions. Some effective interactions, like PKO3-$\Lambda1$, DD-ME2, DDV, and DD-LZ1-$\Lambda1$, exhibit a reduced $R_\Lambda$ with increasing neutron number. Especially, DD-LZ1-$\Lambda1$ gives the smallest hyperon radii among all chosen CDFs, and an strong declining trend. In fact, the core polarization effect due to $\Lambda$ hyperon plays a significant role in this evolution. When $\Lambda$ occupies the $1s_{1/2}$ state, its density distribution is concentrated inside the hypernucleus. As a result, the $\Lambda$'s coupling or attraction with the nucleons in the core (here corresponding to $^{16}$O) appears relatively stronger than that with the valence nucleons. Hence, the evolution of the hyperon radius could be comprehended more or less by the size change of the core with respect to the neutron number.

The variation of the matter radii for the $^{16}$O core in Oxygen (hyper)isotopes is plotted in Fig. \ref{Fig:Rms_tot}(b) with respect to the neutron number. From $N=8$ to 14, in contrast to the situation of total matter radii $R_m$, there is no consistent isospin dependence for the selected CDFs in the core radius $R_m^{\rm{core}}$ with increasing neutron number. The nonlinear RMF functional PK1 exhibits a significant increasing trend with isospin, while the density-dependent RMF one DD-LZ1-$\Lambda$1 shows a noticeable decrease. Consequently, the hyperon radius $R_\Lambda$ exhibit a similar isospin dependence resulting from the core polarization effect, determined mainly by the various isospin properties of CDF functionals in nucleon-nucleon channels. From such analysis, the importance of nuclear in-medium effects in affecting the hyperon radii is unveiled. So the divergent isospin evolution of $R_\Lambda$ given by the CDFs with different density dependent meson-baryon couplings makes it a valuable tool to elucidate the in-medium behavior of nuclear force.

\section{Summary}\label{Summary and Outlook}

In summary, considering the significance of nuclear in-medium effects in nuclear many-body problems, such as eliminating the spurious shell closures, we expanded the newly developed DDRMF Lagrangian DD-LZ1 to incorporate the $\Lambda$ hyperon degree of freedom and determined the $\Lambda N$ effective interaction by fitting the experimental data of $\Lambda$ separation energies for several single-$\Lambda$ hypernuclei. Subsequently, with several other CDF functionals, the features including $\Lambda$ separation energy and $B(E1)$ transition, and the evolution of the spin-orbit splitting as well as the characteristic radii were analyzed in detail along the Oxygen (hyper)isotopes.

By comparing the results obtained from different CDF models, we further investigated the crucial impact of nuclear in-medium effects on accurately describing the properties of hyperon, both in terms of their bulk and single-particle properties. For the $1p$ spin-orbit splitting of the $\Lambda$ hyperon, significant differences in the isospin dependence are observed among the selected CDF effective interactions in Oxygen hyperisotopes. As the neutron number increases, the density circumstance where the hyperon is housing gradually increases, which causes the meson-hyperon coupling strengths that determine the hypernuclear properties to change as well. In particular, the density-dependent CDF effective interactions introduce additional rearrangement terms that significantly enhance the isospin dependence of the $\Lambda$ spin-orbit splitting, leading to more distinct variation of $\Delta E_{\rm{SO}}^\Lambda$ with neutron number in DDRMF and DDRHF models.

The evolution of the hypernuclear matter radius with isospin was further investigated. Significant model dependence in the magnitude of size shrinkage due to the inclusion of $\Lambda$ hyperon is observed, where the DDRMF functional DD-LZ1-$\Lambda$1 displays the largest shrinkage effect. The result was then explained by an anticorrelation between the incompressibility coefficients $K$ of nuclear matter and the hyperon radii $R_\Lambda$, providing us a possible way to constrain the hyperon distribution inside a hypernucleus from better-determined bulk properties of nuclear matter. Additionally, it is found that the isospin evolution of the hyperon radius is primarily influenced by the density-dependent behavior of the chosen CDF functional in $NN$ interaction channel via the procedure of the core polarization. Thus, the sensitivity in depicting these hyperon-relevant properties in CDF models with a various of different meson-baryon couplings holds us great potential to elucidate nuclear in-medium nature in both $\Lambda N$ and $NN$ channels.

\bibliographystyle{apsrev}

\begin{thebibliography}{91}
\expandafter\ifx\csname natexlab\endcsname\relax\def\natexlab#1{#1}\fi
\expandafter\ifx\csname bibnamefont\endcsname\relax
  \def\bibnamefont#1{#1}\fi
\expandafter\ifx\csname bibfnamefont\endcsname\relax
  \def\bibfnamefont#1{#1}\fi
\expandafter\ifx\csname citenamefont\endcsname\relax
  \def\citenamefont#1{#1}\fi
\expandafter\ifx\csname url\endcsname\relax
  \def\url#1{\texttt{#1}}\fi
\expandafter\ifx\csname urlprefix\endcsname\relax\def\urlprefix{URL }\fi
\providecommand{\bibinfo}[2]{#2}
\providecommand{\eprint}[2][]{\url{#2}}

\bibitem[{\citenamefont{Danysz and
  Pniewski}(1953)}]{Danysz1953Philos.Mag.44.348}
\bibinfo{author}{\bibfnamefont{M.}~\bibnamefont{Danysz}} \bibnamefont{and}
  \bibinfo{author}{\bibfnamefont{J.}~\bibnamefont{Pniewski}},
  \bibinfo{journal}{The London, Edinburgh, and Dublin Philosophical Magazine
  and Journal of Science} \textbf{\bibinfo{volume}{44}}, \bibinfo{pages}{348}
  (\bibinfo{year}{1953}), \eprint{https://doi.org/10.1080/14786440308520318},
  \urlprefix\url{https://doi.org/10.1080/14786440308520318}.

\bibitem[{\citenamefont{Hashimoto and Tamura}(2006)}]{Hashimoto2006PPNP57.564}
\bibinfo{author}{\bibfnamefont{O.}~\bibnamefont{Hashimoto}} \bibnamefont{and}
  \bibinfo{author}{\bibfnamefont{H.}~\bibnamefont{Tamura}},
  \bibinfo{journal}{Progress in Particle and Nuclear Physics}
  \textbf{\bibinfo{volume}{57}}, \bibinfo{pages}{564} (\bibinfo{year}{2006}),
  ISSN \bibinfo{issn}{0146-6410},
  \urlprefix\url{https://www.sciencedirect.com/science/article/pii/S0146641005000761}.

\bibitem[{\citenamefont{Gal et~al.}(2016)\citenamefont{Gal, Hungerford, and
  Millener}}]{Gal2016Rev.Mod.Phys.88.035004}
\bibinfo{author}{\bibfnamefont{A.}~\bibnamefont{Gal}},
  \bibinfo{author}{\bibfnamefont{E.~V.} \bibnamefont{Hungerford}},
  \bibnamefont{and} \bibinfo{author}{\bibfnamefont{D.~J.}
  \bibnamefont{Millener}}, \bibinfo{journal}{Rev. Mod. Phys.}
  \textbf{\bibinfo{volume}{88}}, \bibinfo{pages}{035004}
  (\bibinfo{year}{2016}),
  \urlprefix\url{https://link.aps.org/doi/10.1103/RevModPhys.88.035004}.

\bibitem[{\citenamefont{Prakash et~al.}(1997)\citenamefont{Prakash, Bombaci,
  Prakash, Ellis, Lattimer, and Knorren}}]{Prakash9971Phys.Rep.280.1}
\bibinfo{author}{\bibfnamefont{M.}~\bibnamefont{Prakash}},
  \bibinfo{author}{\bibfnamefont{I.}~\bibnamefont{Bombaci}},
  \bibinfo{author}{\bibfnamefont{M.}~\bibnamefont{Prakash}},
  \bibinfo{author}{\bibfnamefont{P.~J.} \bibnamefont{Ellis}},
  \bibinfo{author}{\bibfnamefont{J.~M.} \bibnamefont{Lattimer}},
  \bibnamefont{and} \bibinfo{author}{\bibfnamefont{R.}~\bibnamefont{Knorren}},
  \bibinfo{journal}{Physics Reports} \textbf{\bibinfo{volume}{280}},
  \bibinfo{pages}{1} (\bibinfo{year}{1997}), ISSN \bibinfo{issn}{0370-1573},
  \urlprefix\url{https://www.sciencedirect.com/science/article/pii/S0370157396000233}.

\bibitem[{\citenamefont{Tolos and Fabbietti}(2020)}]{Tolos2020PPNP112.103770}
\bibinfo{author}{\bibfnamefont{L.}~\bibnamefont{Tolos}} \bibnamefont{and}
  \bibinfo{author}{\bibfnamefont{L.}~\bibnamefont{Fabbietti}},
  \bibinfo{journal}{Progress in Particle and Nuclear Physics}
  \textbf{\bibinfo{volume}{112}}, \bibinfo{pages}{103770}
  (\bibinfo{year}{2020}), ISSN \bibinfo{issn}{0146-6410},
  \urlprefix\url{https://www.sciencedirect.com/science/article/pii/S014664102030017X}.

\bibitem[{\citenamefont{Burgio et~al.}(2021)\citenamefont{Burgio, Schulze, na,
  and Wei}}]{Burgio2021PPNP120.103879}
\bibinfo{author}{\bibfnamefont{G.}~\bibnamefont{Burgio}},
  \bibinfo{author}{\bibfnamefont{H.-J.} \bibnamefont{Schulze}},
  \bibinfo{author}{\bibfnamefont{I.~V.} \bibnamefont{na}}, \bibnamefont{and}
  \bibinfo{author}{\bibfnamefont{J.-B.} \bibnamefont{Wei}},
  \bibinfo{journal}{Progress in Particle and Nuclear Physics}
  \textbf{\bibinfo{volume}{120}}, \bibinfo{pages}{103879}
  (\bibinfo{year}{2021}), ISSN \bibinfo{issn}{0146-6410},
  \urlprefix\url{https://www.sciencedirect.com/science/article/pii/S0146641021000338}.

\bibitem[{\citenamefont{Sawada}(2007)}]{Sawada2007NPA782.434}
\bibinfo{author}{\bibfnamefont{S.}~\bibnamefont{Sawada}},
  \bibinfo{journal}{Nuclear Physics A} \textbf{\bibinfo{volume}{782}},
  \bibinfo{pages}{434} (\bibinfo{year}{2007}), ISSN \bibinfo{issn}{0375-9474},
  \urlprefix\url{https://www.sciencedirect.com/science/article/pii/S0375947406007305}.

\bibitem[{\citenamefont{Nakamura et~al.}(2005)\citenamefont{Nakamura,
  Hashimoto, Fujii, Tamura, Takahashi, Maeda, Kanda, Okayasu, Nomura, Matsumura
  et~al.}}]{Nakamura2005NPA754.421}
\bibinfo{author}{\bibfnamefont{S.~N.} \bibnamefont{Nakamura}},
  \bibinfo{author}{\bibfnamefont{O.}~\bibnamefont{Hashimoto}},
  \bibinfo{author}{\bibfnamefont{Y.}~\bibnamefont{Fujii}},
  \bibinfo{author}{\bibfnamefont{H.}~\bibnamefont{Tamura}},
  \bibinfo{author}{\bibfnamefont{T.}~\bibnamefont{Takahashi}},
  \bibinfo{author}{\bibfnamefont{K.}~\bibnamefont{Maeda}},
  \bibinfo{author}{\bibfnamefont{H.}~\bibnamefont{Kanda}},
  \bibinfo{author}{\bibfnamefont{Y.}~\bibnamefont{Okayasu}},
  \bibinfo{author}{\bibfnamefont{H.}~\bibnamefont{Nomura}},
  \bibinfo{author}{\bibfnamefont{A.}~\bibnamefont{Matsumura}},
  \bibnamefont{et~al.}, \bibinfo{journal}{Nuclear Physics A}
  \textbf{\bibinfo{volume}{754}}, \bibinfo{pages}{421} (\bibinfo{year}{2005}),
  ISSN \bibinfo{issn}{0375-9474}, \bibinfo{note}{proceedings of the Eighth
  International Conference on Hypernuclear and Strange Particle Physics},
  \urlprefix\url{https://www.sciencedirect.com/science/article/pii/S037594740500076X}.

\bibitem[{\citenamefont{Henning}(2004)}]{Henning2004NPA734.654}
\bibinfo{author}{\bibfnamefont{W.}~\bibnamefont{Henning}},
  \bibinfo{journal}{Nuclear Physics A} \textbf{\bibinfo{volume}{734}},
  \bibinfo{pages}{654} (\bibinfo{year}{2004}), ISSN \bibinfo{issn}{0375-9474},
  \urlprefix\url{https://www.sciencedirect.com/science/article/pii/S0375947404001393}.

\bibitem[{\citenamefont{Pile et~al.}(1991)\citenamefont{Pile, Bart, Chrien,
  Millener, Sutter, Tsoupas, Peng, Mishra, Hungerford, Kishimoto
  et~al.}}]{Pile1991PRL66.2585}
\bibinfo{author}{\bibfnamefont{P.~H.} \bibnamefont{Pile}},
  \bibinfo{author}{\bibfnamefont{S.}~\bibnamefont{Bart}},
  \bibinfo{author}{\bibfnamefont{R.~E.} \bibnamefont{Chrien}},
  \bibinfo{author}{\bibfnamefont{D.~J.} \bibnamefont{Millener}},
  \bibinfo{author}{\bibfnamefont{R.~J.} \bibnamefont{Sutter}},
  \bibinfo{author}{\bibfnamefont{N.}~\bibnamefont{Tsoupas}},
  \bibinfo{author}{\bibfnamefont{J.-C.} \bibnamefont{Peng}},
  \bibinfo{author}{\bibfnamefont{C.~S.} \bibnamefont{Mishra}},
  \bibinfo{author}{\bibfnamefont{E.~V.} \bibnamefont{Hungerford}},
  \bibinfo{author}{\bibfnamefont{T.}~\bibnamefont{Kishimoto}},
  \bibnamefont{et~al.}, \bibinfo{journal}{Phys. Rev. Lett.}
  \textbf{\bibinfo{volume}{66}}, \bibinfo{pages}{2585} (\bibinfo{year}{1991}),
  \urlprefix\url{https://link.aps.org/doi/10.1103/PhysRevLett.66.2585}.

\bibitem[{\citenamefont{Feliciello and
  Nagae}(2015)}]{Feliciello2015Rep.Prog.Phys.78.096301}
\bibinfo{author}{\bibfnamefont{A.}~\bibnamefont{Feliciello}} \bibnamefont{and}
  \bibinfo{author}{\bibfnamefont{T.}~\bibnamefont{Nagae}},
  \bibinfo{journal}{Reports on Progress in Physics}
  \textbf{\bibinfo{volume}{78}}, \bibinfo{pages}{096301}
  (\bibinfo{year}{2015}),
  \urlprefix\url{https://doi.org/10.1088/0034-4885/78/9/096301}.

\bibitem[{\citenamefont{Tanida et~al.}(2001)\citenamefont{Tanida, Tamura, Abe,
  Akikawa, Araki, Bhang, Endo, Fujii, Fukuda, Hashimoto
  et~al.}}]{Tanida2001PRL86.1982}
\bibinfo{author}{\bibfnamefont{K.}~\bibnamefont{Tanida}},
  \bibinfo{author}{\bibfnamefont{H.}~\bibnamefont{Tamura}},
  \bibinfo{author}{\bibfnamefont{D.}~\bibnamefont{Abe}},
  \bibinfo{author}{\bibfnamefont{H.}~\bibnamefont{Akikawa}},
  \bibinfo{author}{\bibfnamefont{K.}~\bibnamefont{Araki}},
  \bibinfo{author}{\bibfnamefont{H.}~\bibnamefont{Bhang}},
  \bibinfo{author}{\bibfnamefont{T.}~\bibnamefont{Endo}},
  \bibinfo{author}{\bibfnamefont{Y.}~\bibnamefont{Fujii}},
  \bibinfo{author}{\bibfnamefont{T.}~\bibnamefont{Fukuda}},
  \bibinfo{author}{\bibfnamefont{O.}~\bibnamefont{Hashimoto}},
  \bibnamefont{et~al.}, \bibinfo{journal}{Phys. Rev. Lett.}
  \textbf{\bibinfo{volume}{86}}, \bibinfo{pages}{1982} (\bibinfo{year}{2001}),
  \urlprefix\url{https://link.aps.org/doi/10.1103/PhysRevLett.86.1982}.

\bibitem[{\citenamefont{Kohri et~al.}(2002)\citenamefont{Kohri, Ajimura,
  Hayakawa, Kishimoto, Matsuoka, Minami, Miyake, Mori, Morikubo, Saji
  et~al.}}]{Kohri2002PRC65.034607}
\bibinfo{author}{\bibfnamefont{H.}~\bibnamefont{Kohri}},
  \bibinfo{author}{\bibfnamefont{S.}~\bibnamefont{Ajimura}},
  \bibinfo{author}{\bibfnamefont{H.}~\bibnamefont{Hayakawa}},
  \bibinfo{author}{\bibfnamefont{T.}~\bibnamefont{Kishimoto}},
  \bibinfo{author}{\bibfnamefont{K.}~\bibnamefont{Matsuoka}},
  \bibinfo{author}{\bibfnamefont{S.}~\bibnamefont{Minami}},
  \bibinfo{author}{\bibfnamefont{Y.~S.} \bibnamefont{Miyake}},
  \bibinfo{author}{\bibfnamefont{T.}~\bibnamefont{Mori}},
  \bibinfo{author}{\bibfnamefont{K.}~\bibnamefont{Morikubo}},
  \bibinfo{author}{\bibfnamefont{E.}~\bibnamefont{Saji}}, \bibnamefont{et~al.},
  \bibinfo{journal}{Phys. Rev. C} \textbf{\bibinfo{volume}{65}},
  \bibinfo{pages}{034607} (\bibinfo{year}{2002}),
  \urlprefix\url{https://link.aps.org/doi/10.1103/PhysRevC.65.034607}.

\bibitem[{\citenamefont{Feng}(2020)}]{Feng2020PRC102.044604}
\bibinfo{author}{\bibfnamefont{Z.-Q.} \bibnamefont{Feng}},
  \bibinfo{journal}{Phys. Rev. C} \textbf{\bibinfo{volume}{102}},
  \bibinfo{pages}{044604} (\bibinfo{year}{2020}),
  \urlprefix\url{https://link.aps.org/doi/10.1103/PhysRevC.102.044604}.

\bibitem[{\citenamefont{Saito et~al.}(2021)\citenamefont{Saito, Dou, Drozd,
  Ekawa, Escrig, He, Kalantar-Nayestanaki, Kasagi, Kavatsyuk, Liu
  et~al.}}]{Saito2021Nat.Rev.Phys.3.803}
\bibinfo{author}{\bibfnamefont{T.~R.} \bibnamefont{Saito}},
  \bibinfo{author}{\bibfnamefont{W.}~\bibnamefont{Dou}},
  \bibinfo{author}{\bibfnamefont{V.}~\bibnamefont{Drozd}},
  \bibinfo{author}{\bibfnamefont{H.}~\bibnamefont{Ekawa}},
  \bibinfo{author}{\bibfnamefont{S.}~\bibnamefont{Escrig}},
  \bibinfo{author}{\bibfnamefont{Y.}~\bibnamefont{He}},
  \bibinfo{author}{\bibfnamefont{N.}~\bibnamefont{Kalantar-Nayestanaki}},
  \bibinfo{author}{\bibfnamefont{A.}~\bibnamefont{Kasagi}},
  \bibinfo{author}{\bibfnamefont{M.}~\bibnamefont{Kavatsyuk}},
  \bibinfo{author}{\bibfnamefont{E.}~\bibnamefont{Liu}}, \bibnamefont{et~al.},
  \bibinfo{journal}{Nature Reviews Physics} \textbf{\bibinfo{volume}{3}},
  \bibinfo{pages}{803} (\bibinfo{year}{2021}),
  \urlprefix\url{https://doi.org/10.1038/s42254-021-00371-w}.

\bibitem[{\citenamefont{Aboona et~al.}(2023)\citenamefont{Aboona, Adam, Adams,
  Agakishiev, Aggarwal, Aggarwal, Ahammed, Aitbaev, Alekseev, Anderson
  et~al.}}]{Aboona2023PRL130.212301}
\bibinfo{author}{\bibfnamefont{B.~E.} \bibnamefont{Aboona}},
  \bibinfo{author}{\bibfnamefont{J.}~\bibnamefont{Adam}},
  \bibinfo{author}{\bibfnamefont{J.~R.} \bibnamefont{Adams}},
  \bibinfo{author}{\bibfnamefont{G.}~\bibnamefont{Agakishiev}},
  \bibinfo{author}{\bibfnamefont{I.}~\bibnamefont{Aggarwal}},
  \bibinfo{author}{\bibfnamefont{M.~M.} \bibnamefont{Aggarwal}},
  \bibinfo{author}{\bibfnamefont{Z.}~\bibnamefont{Ahammed}},
  \bibinfo{author}{\bibfnamefont{A.}~\bibnamefont{Aitbaev}},
  \bibinfo{author}{\bibfnamefont{I.}~\bibnamefont{Alekseev}},
  \bibinfo{author}{\bibfnamefont{D.~M.} \bibnamefont{Anderson}},
  \bibnamefont{et~al.} (\bibinfo{collaboration}{STAR Collaboration}),
  \bibinfo{journal}{Phys. Rev. Lett.} \textbf{\bibinfo{volume}{130}},
  \bibinfo{pages}{212301} (\bibinfo{year}{2023}),
  \urlprefix\url{https://link.aps.org/doi/10.1103/PhysRevLett.130.212301}.

\bibitem[{\citenamefont{Yang et~al.}(2013)\citenamefont{Yang, Xia, Xiao, Xu,
  Zhao, Zhou, Ma, He, Ma, Gao et~al.}}]{Yang2013NIMPR317.263}
\bibinfo{author}{\bibfnamefont{J.~C.} \bibnamefont{Yang}},
  \bibinfo{author}{\bibfnamefont{J.~W.} \bibnamefont{Xia}},
  \bibinfo{author}{\bibfnamefont{G.~Q.} \bibnamefont{Xiao}},
  \bibinfo{author}{\bibfnamefont{H.~S.} \bibnamefont{Xu}},
  \bibinfo{author}{\bibfnamefont{H.~W.} \bibnamefont{Zhao}},
  \bibinfo{author}{\bibfnamefont{X.~H.} \bibnamefont{Zhou}},
  \bibinfo{author}{\bibfnamefont{X.~W.} \bibnamefont{Ma}},
  \bibinfo{author}{\bibfnamefont{Y.}~\bibnamefont{He}},
  \bibinfo{author}{\bibfnamefont{L.~Z.} \bibnamefont{Ma}},
  \bibinfo{author}{\bibfnamefont{D.~Q.} \bibnamefont{Gao}},
  \bibnamefont{et~al.}, \bibinfo{journal}{Nuclear Instruments and Methods in
  Physics Research Section B: Beam Interactions with Materials and Atoms}
  \textbf{\bibinfo{volume}{317}}, \bibinfo{pages}{263} (\bibinfo{year}{2013}),
  ISSN \bibinfo{issn}{0168-583X},
  \urlprefix\url{https://www.sciencedirect.com/science/article/pii/S0168583X13009877}.

\bibitem[{\citenamefont{Zhou et~al.}(2022)\citenamefont{Zhou, Yang, and {the
  HIAF project team}}}]{Zhou2022AAPPSBulletin32.35}
\bibinfo{author}{\bibfnamefont{X.}~\bibnamefont{Zhou}},
  \bibinfo{author}{\bibfnamefont{J.}~\bibnamefont{Yang}}, \bibnamefont{and}
  \bibinfo{author}{\bibnamefont{{the HIAF project team}}},
  \bibinfo{journal}{AAPPS Bull.} \textbf{\bibinfo{volume}{32}},
  \bibinfo{pages}{35} (\bibinfo{year}{2022}), ISSN \bibinfo{issn}{2309-4710},
  \urlprefix\url{https://link.springer.com/10.1007/s43673-022-00064-1}.

\bibitem[{\citenamefont{Mare\ifmmode~\check{s}\else \v{s}\fi{} and
  Jennings}(1994)}]{Mares1994PRC49.2472}
\bibinfo{author}{\bibfnamefont{J.}~\bibnamefont{Mare\ifmmode~\check{s}\else
  \v{s}\fi{}}} \bibnamefont{and} \bibinfo{author}{\bibfnamefont{B.~K.}
  \bibnamefont{Jennings}}, \bibinfo{journal}{Phys. Rev. C}
  \textbf{\bibinfo{volume}{49}}, \bibinfo{pages}{2472} (\bibinfo{year}{1994}),
  \urlprefix\url{https://link.aps.org/doi/10.1103/PhysRevC.49.2472}.

\bibitem[{\citenamefont{Wirth and Roth}(2018)}]{Wirth2018PLB779.336}
\bibinfo{author}{\bibfnamefont{R.}~\bibnamefont{Wirth}} \bibnamefont{and}
  \bibinfo{author}{\bibfnamefont{R.}~\bibnamefont{Roth}},
  \bibinfo{journal}{Physics Letters B} \textbf{\bibinfo{volume}{779}},
  \bibinfo{pages}{336} (\bibinfo{year}{2018}), ISSN \bibinfo{issn}{0370-2693},
  \urlprefix\url{https://www.sciencedirect.com/science/article/pii/S0370269318301230}.

\bibitem[{\citenamefont{Vretenar et~al.}(1998)\citenamefont{Vretenar, P\"oschl,
  Lalazissis, and Ring}}]{Vretenar1998PRC57.R1060}
\bibinfo{author}{\bibfnamefont{D.}~\bibnamefont{Vretenar}},
  \bibinfo{author}{\bibfnamefont{W.}~\bibnamefont{P\"oschl}},
  \bibinfo{author}{\bibfnamefont{G.~A.} \bibnamefont{Lalazissis}},
  \bibnamefont{and} \bibinfo{author}{\bibfnamefont{P.}~\bibnamefont{Ring}},
  \bibinfo{journal}{Phys. Rev. C} \textbf{\bibinfo{volume}{57}},
  \bibinfo{pages}{R1060} (\bibinfo{year}{1998}),
  \urlprefix\url{https://link.aps.org/doi/10.1103/PhysRevC.57.R1060}.

\bibitem[{\citenamefont{Umeya and Harada}(2011)}]{Umeya2011PRC83.034310}
\bibinfo{author}{\bibfnamefont{A.}~\bibnamefont{Umeya}} \bibnamefont{and}
  \bibinfo{author}{\bibfnamefont{T.}~\bibnamefont{Harada}},
  \bibinfo{journal}{Phys. Rev. C} \textbf{\bibinfo{volume}{83}},
  \bibinfo{pages}{034310} (\bibinfo{year}{2011}),
  \urlprefix\url{https://link.aps.org/doi/10.1103/PhysRevC.83.034310}.

\bibitem[{\citenamefont{Xia et~al.}(2017)\citenamefont{Xia, Mei, and
  Yao}}]{Xia2017Sci.China-Phys.Mech.Astron60.102021}
\bibinfo{author}{\bibfnamefont{H.~J.} \bibnamefont{Xia}},
  \bibinfo{author}{\bibfnamefont{H.}~\bibnamefont{Mei}}, \bibnamefont{and}
  \bibinfo{author}{\bibfnamefont{J.~M.} \bibnamefont{Yao}},
  \bibinfo{journal}{Sci. China-Phys. Mech. Astron}
  \textbf{\bibinfo{volume}{60}}, \bibinfo{pages}{102021}
  (\bibinfo{year}{2017}),
  \urlprefix\url{https://doi.org/10.1007/s11433-017-9048-2}.

\bibitem[{\citenamefont{Ning et~al.}(2009)\citenamefont{Ning, Xian-Rong, and
  Fang-Qi}}]{Wei2009CPC33.116}
\bibinfo{author}{\bibfnamefont{W.}~\bibnamefont{Ning}},
  \bibinfo{author}{\bibfnamefont{Z.}~\bibnamefont{Xian-Rong}},
  \bibnamefont{and} \bibinfo{author}{\bibfnamefont{C.}~\bibnamefont{Fang-Qi}},
  \bibinfo{journal}{Chinese Physics C} \textbf{\bibinfo{volume}{33}},
  \bibinfo{pages}{116} (\bibinfo{year}{2009}),
  \urlprefix\url{https://dx.doi.org/10.1088/1674-1137/33/S1/037}.

\bibitem[{\citenamefont{Lu et~al.}(2011)\citenamefont{Lu, Zhao, and
  Zhou}}]{Lu2011PRC84.014328}
\bibinfo{author}{\bibfnamefont{B.~N.} \bibnamefont{Lu}},
  \bibinfo{author}{\bibfnamefont{E.~G.} \bibnamefont{Zhao}}, \bibnamefont{and}
  \bibinfo{author}{\bibfnamefont{S.~G.} \bibnamefont{Zhou}},
  \bibinfo{journal}{Phys. Rev. C} \textbf{\bibinfo{volume}{84}},
  \bibinfo{pages}{014328} (\bibinfo{year}{2011}),
  \urlprefix\url{https://link.aps.org/doi/10.1103/PhysRevC.84.014328}.

\bibitem[{\citenamefont{Zhang et~al.}(2021)\citenamefont{Zhang, Sagawa, and
  Hiyama}}]{Zhang2021PRC103.034321}
\bibinfo{author}{\bibfnamefont{Y.}~\bibnamefont{Zhang}},
  \bibinfo{author}{\bibfnamefont{H.}~\bibnamefont{Sagawa}}, \bibnamefont{and}
  \bibinfo{author}{\bibfnamefont{E.}~\bibnamefont{Hiyama}},
  \bibinfo{journal}{Phys. Rev. C} \textbf{\bibinfo{volume}{103}},
  \bibinfo{pages}{034321} (\bibinfo{year}{2021}),
  \urlprefix\url{https://link.aps.org/doi/10.1103/PhysRevC.103.034321}.

\bibitem[{\citenamefont{Zhang et~al.}(2022{\natexlab{a}})\citenamefont{Zhang,
  Sagawa, and Hiyama}}]{Zhang2022PTEP2022.023D01}
\bibinfo{author}{\bibfnamefont{Y.}~\bibnamefont{Zhang}},
  \bibinfo{author}{\bibfnamefont{H.}~\bibnamefont{Sagawa}}, \bibnamefont{and}
  \bibinfo{author}{\bibfnamefont{E.}~\bibnamefont{Hiyama}},
  \bibinfo{journal}{Progress of Theoretical and Experimental Physics}
  \textbf{\bibinfo{volume}{2022}} (\bibinfo{year}{2022}{\natexlab{a}}), ISSN
  \bibinfo{issn}{2050-3911}, \bibinfo{note}{023D01},
  \eprint{https://academic.oup.com/ptep/article-pdf/2022/2/023D01/42931223/ptac004.pdf},
  \urlprefix\url{https://doi.org/10.1093/ptep/ptac004}.

\bibitem[{\citenamefont{Xue et~al.}(2022)\citenamefont{Xue, Chen, Zhou, Cheng,
  and Schulze}}]{Xue2022PRC106.044306}
\bibinfo{author}{\bibfnamefont{H.-T.} \bibnamefont{Xue}},
  \bibinfo{author}{\bibfnamefont{Q.~B.} \bibnamefont{Chen}},
  \bibinfo{author}{\bibfnamefont{X.-R.} \bibnamefont{Zhou}},
  \bibinfo{author}{\bibfnamefont{Y.~Y.} \bibnamefont{Cheng}}, \bibnamefont{and}
  \bibinfo{author}{\bibfnamefont{H.-J.} \bibnamefont{Schulze}},
  \bibinfo{journal}{Phys. Rev. C} \textbf{\bibinfo{volume}{106}},
  \bibinfo{pages}{044306} (\bibinfo{year}{2022}),
  \urlprefix\url{https://link.aps.org/doi/10.1103/PhysRevC.106.044306}.

\bibitem[{\citenamefont{Reinhard}(1989)}]{Reinhard1989Rep.Prog.Phys.52.439}
\bibinfo{author}{\bibfnamefont{P.~G.} \bibnamefont{Reinhard}},
  \bibinfo{journal}{Reports on Progress in Physics}
  \textbf{\bibinfo{volume}{52}}, \bibinfo{pages}{439} (\bibinfo{year}{1989}),
  \urlprefix\url{https://doi.org/10.1088/0034-4885/52/4/002}.

\bibitem[{\citenamefont{Ring}(1996)}]{Ring1996PPNP37.193}
\bibinfo{author}{\bibfnamefont{P.}~\bibnamefont{Ring}},
  \bibinfo{journal}{Progress in Particle and Nuclear Physics}
  \textbf{\bibinfo{volume}{37}}, \bibinfo{pages}{193} (\bibinfo{year}{1996}),
  ISSN \bibinfo{issn}{0146-6410},
  \urlprefix\url{https://www.sciencedirect.com/science/article/pii/0146641096000543}.

\bibitem[{\citenamefont{Bender et~al.}(2003)\citenamefont{Bender, Heenen, and
  Reinhard}}]{Bender2003Rev.Mod.Phys.75.121}
\bibinfo{author}{\bibfnamefont{M.}~\bibnamefont{Bender}},
  \bibinfo{author}{\bibfnamefont{P.~H.} \bibnamefont{Heenen}},
  \bibnamefont{and} \bibinfo{author}{\bibfnamefont{P.~G.}
  \bibnamefont{Reinhard}}, \bibinfo{journal}{Rev. Mod. Phys.}
  \textbf{\bibinfo{volume}{75}}, \bibinfo{pages}{121} (\bibinfo{year}{2003}),
  \urlprefix\url{https://link.aps.org/doi/10.1103/RevModPhys.75.121}.

\bibitem[{\citenamefont{Vretenar et~al.}(2005)\citenamefont{Vretenar,
  Afanasjev, Lalazissis, and Ring}}]{Vretenar2005Phys.Rep.409.101}
\bibinfo{author}{\bibfnamefont{D.}~\bibnamefont{Vretenar}},
  \bibinfo{author}{\bibfnamefont{A.~V.} \bibnamefont{Afanasjev}},
  \bibinfo{author}{\bibfnamefont{G.~A.} \bibnamefont{Lalazissis}},
  \bibnamefont{and} \bibinfo{author}{\bibfnamefont{P.}~\bibnamefont{Ring}},
  \bibinfo{journal}{Physics Reports} \textbf{\bibinfo{volume}{409}},
  \bibinfo{pages}{101} (\bibinfo{year}{2005}), ISSN \bibinfo{issn}{0370-1573},
  \urlprefix\url{https://www.sciencedirect.com/science/article/pii/S0370157304004545}.

\bibitem[{\citenamefont{Meng et~al.}(2006)\citenamefont{Meng, Toki, Zhou,
  Zhang, Long, and Geng}}]{Meng2006PPNP57.470}
\bibinfo{author}{\bibfnamefont{J.}~\bibnamefont{Meng}},
  \bibinfo{author}{\bibfnamefont{H.}~\bibnamefont{Toki}},
  \bibinfo{author}{\bibfnamefont{S.~G.} \bibnamefont{Zhou}},
  \bibinfo{author}{\bibfnamefont{S.~Q.} \bibnamefont{Zhang}},
  \bibinfo{author}{\bibfnamefont{W.~H.} \bibnamefont{Long}}, \bibnamefont{and}
  \bibinfo{author}{\bibfnamefont{L.~S.} \bibnamefont{Geng}},
  \bibinfo{journal}{Progress in Particle and Nuclear Physics}
  \textbf{\bibinfo{volume}{57}}, \bibinfo{pages}{470} (\bibinfo{year}{2006}),
  ISSN \bibinfo{issn}{0146-6410},
  \urlprefix\url{https://www.sciencedirect.com/science/article/pii/S014664100500075X}.

\bibitem[{\citenamefont{Nik\v{s}i\'{c}
  et~al.}(2011)\citenamefont{Nik\v{s}i\'{c}, Vretenar, and
  Ring}}]{Niksic2011PPNP66.519}
\bibinfo{author}{\bibfnamefont{T.}~\bibnamefont{Nik\v{s}i\'{c}}},
  \bibinfo{author}{\bibfnamefont{D.}~\bibnamefont{Vretenar}}, \bibnamefont{and}
  \bibinfo{author}{\bibfnamefont{P.}~\bibnamefont{Ring}},
  \bibinfo{journal}{Progress in Particle and Nuclear Physics}
  \textbf{\bibinfo{volume}{66}}, \bibinfo{pages}{519} (\bibinfo{year}{2011}),
  ISSN \bibinfo{issn}{0146-6410},
  \urlprefix\url{https://www.sciencedirect.com/science/article/pii/S0146641011000561}.

\bibitem[{\citenamefont{Meng and Zhou}(2015)}]{Meng2015JPG42.093101}
\bibinfo{author}{\bibfnamefont{J.}~\bibnamefont{Meng}} \bibnamefont{and}
  \bibinfo{author}{\bibfnamefont{S.~G.} \bibnamefont{Zhou}},
  \bibinfo{journal}{Journal of Physics G: Nuclear and Particle Physics}
  \textbf{\bibinfo{volume}{42}}, \bibinfo{pages}{093101}
  (\bibinfo{year}{2015}),
  \urlprefix\url{https://doi.org/10.1088/0954-3899/42/9/093101}.

\bibitem[{\citenamefont{Meng}(2016)}]{Meng2016Density}
\bibinfo{author}{\bibfnamefont{J.}~\bibnamefont{Meng}},
  \emph{\bibinfo{title}{Relativistic Density Functional for Nuclear Structure}}
  (\bibinfo{publisher}{WORLD SCIENTIFIC}, \bibinfo{year}{2016}),
  \eprint{https://www.worldscientific.com/doi/pdf/10.1142/9872},
  \urlprefix\url{https://www.worldscientific.com/doi/abs/10.1142/9872}.

\bibitem[{\citenamefont{Rayet}(1976)}]{Rayet1976Ann.Phys.102.226}
\bibinfo{author}{\bibfnamefont{M.}~\bibnamefont{Rayet}},
  \bibinfo{journal}{Annals of Physics} \textbf{\bibinfo{volume}{102}},
  \bibinfo{pages}{226} (\bibinfo{year}{1976}), ISSN \bibinfo{issn}{0003-4916},
  \urlprefix\url{https://www.sciencedirect.com/science/article/pii/0003491676902621}.

\bibitem[{\citenamefont{Lanskoy and Yamamoto}(1997)}]{Lanskoy1997PRC55.2330}
\bibinfo{author}{\bibfnamefont{D.~E.} \bibnamefont{Lanskoy}} \bibnamefont{and}
  \bibinfo{author}{\bibfnamefont{Y.}~\bibnamefont{Yamamoto}},
  \bibinfo{journal}{Phys. Rev. C} \textbf{\bibinfo{volume}{55}},
  \bibinfo{pages}{2330} (\bibinfo{year}{1997}),
  \urlprefix\url{https://link.aps.org/doi/10.1103/PhysRevC.55.2330}.

\bibitem[{\citenamefont{Brockmann and Weise}(1977)}]{Brockmann1977PLB69.167}
\bibinfo{author}{\bibfnamefont{R.}~\bibnamefont{Brockmann}} \bibnamefont{and}
  \bibinfo{author}{\bibfnamefont{W.}~\bibnamefont{Weise}},
  \bibinfo{journal}{Physics Letters B} \textbf{\bibinfo{volume}{69}},
  \bibinfo{pages}{167} (\bibinfo{year}{1977}), ISSN \bibinfo{issn}{0370-2693},
  \urlprefix\url{https://www.sciencedirect.com/science/article/pii/0370269377906359}.

\bibitem[{\citenamefont{Bouyssy}(1981)}]{Bouyssy1981PLB99.305}
\bibinfo{author}{\bibfnamefont{A.}~\bibnamefont{Bouyssy}},
  \bibinfo{journal}{Physics Letters B} \textbf{\bibinfo{volume}{99}},
  \bibinfo{pages}{305} (\bibinfo{year}{1981}), ISSN \bibinfo{issn}{0370-2693},
  \urlprefix\url{https://www.sciencedirect.com/science/article/pii/0370269381901064}.

\bibitem[{\citenamefont{Glendenning and
  Moszkowski}(1991)}]{Glendenning1991PRL67.2414}
\bibinfo{author}{\bibfnamefont{N.~K.} \bibnamefont{Glendenning}}
  \bibnamefont{and} \bibinfo{author}{\bibfnamefont{S.~A.}
  \bibnamefont{Moszkowski}}, \bibinfo{journal}{Phys. Rev. Lett.}
  \textbf{\bibinfo{volume}{67}}, \bibinfo{pages}{2414} (\bibinfo{year}{1991}),
  \urlprefix\url{https://link.aps.org/doi/10.1103/PhysRevLett.67.2414}.

\bibitem[{\citenamefont{Sugahara and Toki}(1994)}]{Sugahara1994PTP92.803}
\bibinfo{author}{\bibfnamefont{Y.}~\bibnamefont{Sugahara}} \bibnamefont{and}
  \bibinfo{author}{\bibfnamefont{H.}~\bibnamefont{Toki}},
  \bibinfo{journal}{Progress of Theoretical Physics}
  \textbf{\bibinfo{volume}{92}}, \bibinfo{pages}{803} (\bibinfo{year}{1994}),
  ISSN \bibinfo{issn}{0033-068X},
  \eprint{https://academic.oup.com/ptp/article-pdf/92/4/803/5358491/92-4-803.pdf},
  \urlprefix\url{https://doi.org/10.1143/ptp/92.4.803}.

\bibitem[{\citenamefont{Zhou et~al.}(2008)\citenamefont{Zhou, Polls, Schulze,
  and Vida\~na}}]{Zhou2008PRC78.054306}
\bibinfo{author}{\bibfnamefont{X.-R.} \bibnamefont{Zhou}},
  \bibinfo{author}{\bibfnamefont{A.}~\bibnamefont{Polls}},
  \bibinfo{author}{\bibfnamefont{H.-J.} \bibnamefont{Schulze}},
  \bibnamefont{and} \bibinfo{author}{\bibfnamefont{I.}~\bibnamefont{Vida\~na}},
  \bibinfo{journal}{Phys. Rev. C} \textbf{\bibinfo{volume}{78}},
  \bibinfo{pages}{054306} (\bibinfo{year}{2008}),
  \urlprefix\url{https://link.aps.org/doi/10.1103/PhysRevC.78.054306}.

\bibitem[{\citenamefont{Hu et~al.}(2014)\citenamefont{Hu, Hiyama, and
  Toki}}]{Hu2014PRC90.014309}
\bibinfo{author}{\bibfnamefont{J.~N.} \bibnamefont{Hu}},
  \bibinfo{author}{\bibfnamefont{E.}~\bibnamefont{Hiyama}}, \bibnamefont{and}
  \bibinfo{author}{\bibfnamefont{H.}~\bibnamefont{Toki}},
  \bibinfo{journal}{Phys. Rev. C} \textbf{\bibinfo{volume}{90}},
  \bibinfo{pages}{014309} (\bibinfo{year}{2014}),
  \urlprefix\url{https://link.aps.org/doi/10.1103/PhysRevC.90.014309}.

\bibitem[{\citenamefont{Li et~al.}(2018)\citenamefont{Li, Long, and
  Sedrakian}}]{Li2018EPJA54.133}
\bibinfo{author}{\bibfnamefont{J.~J.} \bibnamefont{Li}},
  \bibinfo{author}{\bibfnamefont{W.~H.} \bibnamefont{Long}}, \bibnamefont{and}
  \bibinfo{author}{\bibfnamefont{A.}~\bibnamefont{Sedrakian}},
  \bibinfo{journal}{The European Physical Journal A}
  \textbf{\bibinfo{volume}{54}}, \bibinfo{pages}{133} (\bibinfo{year}{2018}),
  \urlprefix\url{https://doi.org/10.1140/epja/i2018-12566-6}.

\bibitem[{\citenamefont{Rong et~al.}(2020)\citenamefont{Rong, Zhao, and
  Zhou}}]{Zhou2020PLB807.135533}
\bibinfo{author}{\bibfnamefont{Y.~T.} \bibnamefont{Rong}},
  \bibinfo{author}{\bibfnamefont{P.~W.} \bibnamefont{Zhao}}, \bibnamefont{and}
  \bibinfo{author}{\bibfnamefont{S.~G.} \bibnamefont{Zhou}},
  \bibinfo{journal}{Physics Letters B} \textbf{\bibinfo{volume}{807}},
  \bibinfo{pages}{135533} (\bibinfo{year}{2020}), ISSN
  \bibinfo{issn}{0370-2693},
  \urlprefix\url{https://www.sciencedirect.com/science/article/pii/S0370269320303373}.

\bibitem[{\citenamefont{Wu et~al.}(2017)\citenamefont{Wu, Mei, Yao, and
  Zhou}}]{Yao2017PRC95.034309}
\bibinfo{author}{\bibfnamefont{X.~Y.} \bibnamefont{Wu}},
  \bibinfo{author}{\bibfnamefont{H.}~\bibnamefont{Mei}},
  \bibinfo{author}{\bibfnamefont{J.~M.} \bibnamefont{Yao}}, \bibnamefont{and}
  \bibinfo{author}{\bibfnamefont{X.~R.} \bibnamefont{Zhou}},
  \bibinfo{journal}{Phys. Rev. C} \textbf{\bibinfo{volume}{95}},
  \bibinfo{pages}{034309} (\bibinfo{year}{2017}),
  \urlprefix\url{https://link.aps.org/doi/10.1103/PhysRevC.95.034309}.

\bibitem[{\citenamefont{Tanimura and Hagino}(2012)}]{Tanimura2012PRC85.014306}
\bibinfo{author}{\bibfnamefont{Y.}~\bibnamefont{Tanimura}} \bibnamefont{and}
  \bibinfo{author}{\bibfnamefont{K.}~\bibnamefont{Hagino}},
  \bibinfo{journal}{Phys. Rev. C} \textbf{\bibinfo{volume}{85}},
  \bibinfo{pages}{014306} (\bibinfo{year}{2012}),
  \urlprefix\url{https://link.aps.org/doi/10.1103/PhysRevC.85.014306}.

\bibitem[{\citenamefont{Hong-Feng and Jie}(2002)}]{Lv2002CPL19.1775}
\bibinfo{author}{\bibfnamefont{L.}~\bibnamefont{Hong-Feng}} \bibnamefont{and}
  \bibinfo{author}{\bibfnamefont{M.}~\bibnamefont{Jie}},
  \bibinfo{journal}{Chinese Physics Letters} \textbf{\bibinfo{volume}{19}},
  \bibinfo{pages}{1775} (\bibinfo{year}{2002}),
  \urlprefix\url{https://dx.doi.org/10.1088/0256-307X/19/12/310}.

\bibitem[{\citenamefont{Win and Hagino}(2008)}]{Win2008PRC78.054311}
\bibinfo{author}{\bibfnamefont{M.~T.} \bibnamefont{Win}} \bibnamefont{and}
  \bibinfo{author}{\bibfnamefont{K.}~\bibnamefont{Hagino}},
  \bibinfo{journal}{Phys. Rev. C} \textbf{\bibinfo{volume}{78}},
  \bibinfo{pages}{054311} (\bibinfo{year}{2008}),
  \urlprefix\url{https://link.aps.org/doi/10.1103/PhysRevC.78.054311}.

\bibitem[{\citenamefont{Lu et~al.}(2014)\citenamefont{Lu, Hiyama, Sagawa, and
  Zhou}}]{Zhou2014PRC89.044307}
\bibinfo{author}{\bibfnamefont{B.~N.} \bibnamefont{Lu}},
  \bibinfo{author}{\bibfnamefont{E.}~\bibnamefont{Hiyama}},
  \bibinfo{author}{\bibfnamefont{H.}~\bibnamefont{Sagawa}}, \bibnamefont{and}
  \bibinfo{author}{\bibfnamefont{S.~G.} \bibnamefont{Zhou}},
  \bibinfo{journal}{Phys. Rev. C} \textbf{\bibinfo{volume}{89}},
  \bibinfo{pages}{044307} (\bibinfo{year}{2014}),
  \urlprefix\url{https://link.aps.org/doi/10.1103/PhysRevC.89.044307}.

\bibitem[{\citenamefont{Chen et~al.}(2022)\citenamefont{Chen, Chen, Zhou,
  Cheng, Cui, and Schulze}}]{Chen2022CPC46.064109}
\bibinfo{author}{\bibfnamefont{C.~F.} \bibnamefont{Chen}},
  \bibinfo{author}{\bibfnamefont{Q.~B.} \bibnamefont{Chen}},
  \bibinfo{author}{\bibfnamefont{X.-R.} \bibnamefont{Zhou}},
  \bibinfo{author}{\bibfnamefont{Y.~Y.} \bibnamefont{Cheng}},
  \bibinfo{author}{\bibfnamefont{J.-W.} \bibnamefont{Cui}}, \bibnamefont{and}
  \bibinfo{author}{\bibfnamefont{H.-J.} \bibnamefont{Schulze}},
  \bibinfo{journal}{Chinese Physics C} \textbf{\bibinfo{volume}{46}},
  \bibinfo{pages}{064109} (\bibinfo{year}{2022}),
  \urlprefix\url{https://dx.doi.org/10.1088/1674-1137/ac5b58}.

\bibitem[{\citenamefont{Tanimura}(2019)}]{Tanimura2019PRC99.034324}
\bibinfo{author}{\bibfnamefont{Y.}~\bibnamefont{Tanimura}},
  \bibinfo{journal}{Phys. Rev. C} \textbf{\bibinfo{volume}{99}},
  \bibinfo{pages}{034324} (\bibinfo{year}{2019}),
  \urlprefix\url{https://link.aps.org/doi/10.1103/PhysRevC.99.034324}.

\bibitem[{\citenamefont{Meng et~al.}(2003)\citenamefont{Meng, L\"{u}, Zhang,
  and Zhou}}]{Meng2003NPA722.C366}
\bibinfo{author}{\bibfnamefont{J.}~\bibnamefont{Meng}},
  \bibinfo{author}{\bibfnamefont{H.}~\bibnamefont{L\"{u}}},
  \bibinfo{author}{\bibfnamefont{S.}~\bibnamefont{Zhang}}, \bibnamefont{and}
  \bibinfo{author}{\bibfnamefont{S.-G.} \bibnamefont{Zhou}},
  \bibinfo{journal}{Nuclear Physics A} \textbf{\bibinfo{volume}{722}},
  \bibinfo{pages}{C366} (\bibinfo{year}{2003}), ISSN \bibinfo{issn}{0375-9474},
  \urlprefix\url{https://www.sciencedirect.com/science/article/pii/S0375947403013915}.

\bibitem[{\citenamefont{Rong et~al.}(2021)\citenamefont{Rong, Tu, and
  Zhou}}]{Rong2021PRC104.054321}
\bibinfo{author}{\bibfnamefont{Y.-T.} \bibnamefont{Rong}},
  \bibinfo{author}{\bibfnamefont{Z.-H.} \bibnamefont{Tu}}, \bibnamefont{and}
  \bibinfo{author}{\bibfnamefont{S.-G.} \bibnamefont{Zhou}},
  \bibinfo{journal}{Phys. Rev. C} \textbf{\bibinfo{volume}{104}},
  \bibinfo{pages}{054321} (\bibinfo{year}{2021}),
  \urlprefix\url{https://link.aps.org/doi/10.1103/PhysRevC.104.054321}.

\bibitem[{\citenamefont{Wei et~al.}(2020)\citenamefont{Wei, Zhao, Wang, Geng,
  Sun, Niu, and Long}}]{Wei2020CPC44.074107}
\bibinfo{author}{\bibfnamefont{B.}~\bibnamefont{Wei}},
  \bibinfo{author}{\bibfnamefont{Q.}~\bibnamefont{Zhao}},
  \bibinfo{author}{\bibfnamefont{Z.~H.} \bibnamefont{Wang}},
  \bibinfo{author}{\bibfnamefont{J.}~\bibnamefont{Geng}},
  \bibinfo{author}{\bibfnamefont{B.~Y.} \bibnamefont{Sun}},
  \bibinfo{author}{\bibfnamefont{Y.~F.} \bibnamefont{Niu}}, \bibnamefont{and}
  \bibinfo{author}{\bibfnamefont{W.~H.} \bibnamefont{Long}},
  \bibinfo{journal}{Chinese Physics C} \textbf{\bibinfo{volume}{44}},
  \bibinfo{pages}{074107} (\bibinfo{year}{2020}),
  \urlprefix\url{https://doi.org/10.1088/1674-1137/44/7/074107}.

\bibitem[{\citenamefont{Zhang et~al.}(2022{\natexlab{b}})\citenamefont{Zhang,
  Li, Gao, and Sun}}]{Zhang2022CPC46.104.105}
\bibinfo{author}{\bibfnamefont{W.}~\bibnamefont{Zhang}},
  \bibinfo{author}{\bibfnamefont{Z.~Y.} \bibnamefont{Li}},
  \bibinfo{author}{\bibfnamefont{W.}~\bibnamefont{Gao}}, \bibnamefont{and}
  \bibinfo{author}{\bibfnamefont{T.~T.} \bibnamefont{Sun}},
  \bibinfo{journal}{Chinese Physics C} \textbf{\bibinfo{volume}{46}},
  \bibinfo{pages}{104105} (\bibinfo{year}{2022}{\natexlab{b}}),
  \urlprefix\url{https://dx.doi.org/10.1088/1674-1137/ac7b18}.

\bibitem[{\citenamefont{Rather et~al.}(2021{\natexlab{a}})\citenamefont{Rather,
  Rahaman, Dexheimer, Usmani, and Patra}}]{Rather2021APJ917.46}
\bibinfo{author}{\bibfnamefont{I.~A.} \bibnamefont{Rather}},
  \bibinfo{author}{\bibfnamefont{U.}~\bibnamefont{Rahaman}},
  \bibinfo{author}{\bibfnamefont{V.}~\bibnamefont{Dexheimer}},
  \bibinfo{author}{\bibfnamefont{A.~A.} \bibnamefont{Usmani}},
  \bibnamefont{and} \bibinfo{author}{\bibfnamefont{S.~K.} \bibnamefont{Patra}},
  \bibinfo{journal}{The Astrophysical Journal} \textbf{\bibinfo{volume}{917}},
  \bibinfo{pages}{46} (\bibinfo{year}{2021}{\natexlab{a}}),
  \urlprefix\url{https://dx.doi.org/10.3847/1538-4357/ac09f7}.

\bibitem[{\citenamefont{Sun et~al.}(2023)\citenamefont{Sun, Miao, Sun, and
  Li}}]{Sun2023APJ942.55}
\bibinfo{author}{\bibfnamefont{X.}~\bibnamefont{Sun}},
  \bibinfo{author}{\bibfnamefont{Z.}~\bibnamefont{Miao}},
  \bibinfo{author}{\bibfnamefont{B.}~\bibnamefont{Sun}}, \bibnamefont{and}
  \bibinfo{author}{\bibfnamefont{A.}~\bibnamefont{Li}}, \bibinfo{journal}{The
  Astrophysical Journal} \textbf{\bibinfo{volume}{942}}, \bibinfo{pages}{55}
  (\bibinfo{year}{2023}),
  \urlprefix\url{https://dx.doi.org/10.3847/1538-4357/ac9d9a}.

\bibitem[{\citenamefont{Rather et~al.}(2021{\natexlab{b}})\citenamefont{Rather,
  Rahaman, Imran, Das, Usmani, and Patra}}]{Rather2021PRC103.055814}
\bibinfo{author}{\bibfnamefont{I.~A.} \bibnamefont{Rather}},
  \bibinfo{author}{\bibfnamefont{U.}~\bibnamefont{Rahaman}},
  \bibinfo{author}{\bibfnamefont{M.}~\bibnamefont{Imran}},
  \bibinfo{author}{\bibfnamefont{H.~C.} \bibnamefont{Das}},
  \bibinfo{author}{\bibfnamefont{A.~A.} \bibnamefont{Usmani}},
  \bibnamefont{and} \bibinfo{author}{\bibfnamefont{S.~K.} \bibnamefont{Patra}},
  \bibinfo{journal}{Phys. Rev. C} \textbf{\bibinfo{volume}{103}},
  \bibinfo{pages}{055814} (\bibinfo{year}{2021}{\natexlab{b}}),
  \urlprefix\url{https://link.aps.org/doi/10.1103/PhysRevC.103.055814}.

\bibitem[{\citenamefont{Malik et~al.}(2022)\citenamefont{Malik, Ferreira,
  Agrawal, and Provid\^encia}}]{Malik2022APJ930.17}
\bibinfo{author}{\bibfnamefont{T.}~\bibnamefont{Malik}},
  \bibinfo{author}{\bibfnamefont{M.}~\bibnamefont{Ferreira}},
  \bibinfo{author}{\bibfnamefont{B.~K.} \bibnamefont{Agrawal}},
  \bibnamefont{and}
  \bibinfo{author}{\bibfnamefont{C.}~\bibnamefont{Provid\^encia}},
  \bibinfo{journal}{The Astrophysical Journal} \textbf{\bibinfo{volume}{930}},
  \bibinfo{pages}{17} (\bibinfo{year}{2022}),
  \urlprefix\url{https://dx.doi.org/10.3847/1538-4357/ac5d3c}.

\bibitem[{\citenamefont{Yang et~al.}(2022)\citenamefont{Yang, Wen, Wang, and
  Zhang}}]{Yang2022PRD105.063023}
\bibinfo{author}{\bibfnamefont{S.}~\bibnamefont{Yang}},
  \bibinfo{author}{\bibfnamefont{D.}~\bibnamefont{Wen}},
  \bibinfo{author}{\bibfnamefont{J.}~\bibnamefont{Wang}}, \bibnamefont{and}
  \bibinfo{author}{\bibfnamefont{J.}~\bibnamefont{Zhang}},
  \bibinfo{journal}{Phys. Rev. D} \textbf{\bibinfo{volume}{105}},
  \bibinfo{pages}{063023} (\bibinfo{year}{2022}),
  \urlprefix\url{https://link.aps.org/doi/10.1103/PhysRevD.105.063023}.

\bibitem[{\citenamefont{Xia et~al.}(2022{\natexlab{a}})\citenamefont{Xia, Sun,
  Maruyama, Long, and Li}}]{Xia2022PRC105.045803}
\bibinfo{author}{\bibfnamefont{C.-J.} \bibnamefont{Xia}},
  \bibinfo{author}{\bibfnamefont{B.~Y.} \bibnamefont{Sun}},
  \bibinfo{author}{\bibfnamefont{T.}~\bibnamefont{Maruyama}},
  \bibinfo{author}{\bibfnamefont{W.-H.} \bibnamefont{Long}}, \bibnamefont{and}
  \bibinfo{author}{\bibfnamefont{A.}~\bibnamefont{Li}}, \bibinfo{journal}{Phys.
  Rev. C} \textbf{\bibinfo{volume}{105}}, \bibinfo{pages}{045803}
  (\bibinfo{year}{2022}{\natexlab{a}}),
  \urlprefix\url{https://link.aps.org/doi/10.1103/PhysRevC.105.045803}.

\bibitem[{\citenamefont{Xia et~al.}(2022{\natexlab{b}})\citenamefont{Xia,
  Maruyama, Li, Sun, Long, and Zhang}}]{Xia2022CTP74.095303}
\bibinfo{author}{\bibfnamefont{C.-J.} \bibnamefont{Xia}},
  \bibinfo{author}{\bibfnamefont{T.}~\bibnamefont{Maruyama}},
  \bibinfo{author}{\bibfnamefont{A.}~\bibnamefont{Li}},
  \bibinfo{author}{\bibfnamefont{B.~Y.} \bibnamefont{Sun}},
  \bibinfo{author}{\bibfnamefont{W.-H.} \bibnamefont{Long}}, \bibnamefont{and}
  \bibinfo{author}{\bibfnamefont{Y.-X.} \bibnamefont{Zhang}},
  \bibinfo{journal}{Communications in Theoretical Physics}
  \textbf{\bibinfo{volume}{74}}, \bibinfo{pages}{095303}
  (\bibinfo{year}{2022}{\natexlab{b}}),
  \urlprefix\url{https://dx.doi.org/10.1088/1572-9494/ac71fd}.

\bibitem[{\citenamefont{Isaka et~al.}(2013)\citenamefont{Isaka, Homma, Kimura,
  Dote, and Ohnishi}}]{Isaka2013Few-Body-Systems54.1219}
\bibinfo{author}{\bibfnamefont{M.}~\bibnamefont{Isaka}},
  \bibinfo{author}{\bibfnamefont{H.}~\bibnamefont{Homma}},
  \bibinfo{author}{\bibfnamefont{M.}~\bibnamefont{Kimura}},
  \bibinfo{author}{\bibfnamefont{A.}~\bibnamefont{Dote}}, \bibnamefont{and}
  \bibinfo{author}{\bibfnamefont{A.}~\bibnamefont{Ohnishi}},
  \bibinfo{journal}{Few-Body Systems} \textbf{\bibinfo{volume}{54}},
  \bibinfo{pages}{1219} (\bibinfo{year}{2013}), ISSN \bibinfo{issn}{0177-7963,
  1432-5411},
  \urlprefix\url{http://link.springer.com/10.1007/s00601-012-0547-3}.

\bibitem[{\citenamefont{Choi et~al.}(2022)\citenamefont{Choi, Hiyama, Hyun, and
  Cheoun}}]{Choi2022EPJA58.161}
\bibinfo{author}{\bibfnamefont{S.}~\bibnamefont{Choi}},
  \bibinfo{author}{\bibfnamefont{E.}~\bibnamefont{Hiyama}},
  \bibinfo{author}{\bibfnamefont{C.~H.} \bibnamefont{Hyun}}, \bibnamefont{and}
  \bibinfo{author}{\bibfnamefont{M.-K.} \bibnamefont{Cheoun}},
  \bibinfo{journal}{The European Physical Journal A}
  \textbf{\bibinfo{volume}{58}}, \bibinfo{pages}{161} (\bibinfo{year}{2022}),
  ISSN \bibinfo{issn}{8},
  \urlprefix\url{https://doi.org/10.1140/epja/s10050-022-00817-4}.

\bibitem[{\citenamefont{Aoki et~al.}(2021)\citenamefont{Aoki, Fujioka, Gogami,
  Hidaka, Hiyama, Honda, Hosaka, Ichikawa, Ieiri, Isaka
  et~al.}}]{Aoki2021arXive2110.04462}
\bibinfo{author}{\bibfnamefont{K.}~\bibnamefont{Aoki}},
  \bibinfo{author}{\bibfnamefont{H.}~\bibnamefont{Fujioka}},
  \bibinfo{author}{\bibfnamefont{T.}~\bibnamefont{Gogami}},
  \bibinfo{author}{\bibfnamefont{Y.}~\bibnamefont{Hidaka}},
  \bibinfo{author}{\bibfnamefont{E.}~\bibnamefont{Hiyama}},
  \bibinfo{author}{\bibfnamefont{R.}~\bibnamefont{Honda}},
  \bibinfo{author}{\bibfnamefont{A.}~\bibnamefont{Hosaka}},
  \bibinfo{author}{\bibfnamefont{Y.}~\bibnamefont{Ichikawa}},
  \bibinfo{author}{\bibfnamefont{M.}~\bibnamefont{Ieiri}},
  \bibinfo{author}{\bibfnamefont{M.}~\bibnamefont{Isaka}},
  \bibnamefont{et~al.}, \emph{\bibinfo{title}{Extension of the j-parc hadron
  experimental facility: Third white paper}} (\bibinfo{year}{2021}),
  \eprint{2110.04462}.

\bibitem[{\citenamefont{Long et~al.}(2006)\citenamefont{Long, {Van Giai}, and
  Meng}}]{Long2006PLB640.150}
\bibinfo{author}{\bibfnamefont{W.~H.} \bibnamefont{Long}},
  \bibinfo{author}{\bibfnamefont{N.}~\bibnamefont{{Van Giai}}},
  \bibnamefont{and} \bibinfo{author}{\bibfnamefont{J.}~\bibnamefont{Meng}},
  \bibinfo{journal}{Physics Letters B} \textbf{\bibinfo{volume}{640}},
  \bibinfo{pages}{150} (\bibinfo{year}{2006}), ISSN \bibinfo{issn}{0370-2693},
  \urlprefix\url{https://www.sciencedirect.com/science/article/pii/S0370269306009610}.

\bibitem[{\citenamefont{Ding et~al.}(2022)\citenamefont{Ding, Qian, Sun, and
  Long}}]{Ding2022PRC106.054311}
\bibinfo{author}{\bibfnamefont{S.~Y.} \bibnamefont{Ding}},
  \bibinfo{author}{\bibfnamefont{Z.}~\bibnamefont{Qian}},
  \bibinfo{author}{\bibfnamefont{B.~Y.} \bibnamefont{Sun}}, \bibnamefont{and}
  \bibinfo{author}{\bibfnamefont{W.~H.} \bibnamefont{Long}},
  \bibinfo{journal}{Phys. Rev. C} \textbf{\bibinfo{volume}{106}},
  \bibinfo{pages}{054311} (\bibinfo{year}{2022}),
  \urlprefix\url{https://link.aps.org/doi/10.1103/PhysRevC.106.054311}.

\bibitem[{\citenamefont{Xia et~al.}(2023)\citenamefont{Xia, Wu, Mei, and
  Yao}}]{Xia2023Sci.China-Phys.Mech.Astron66.252011}
\bibinfo{author}{\bibfnamefont{H.}~\bibnamefont{Xia}},
  \bibinfo{author}{\bibfnamefont{X.}~\bibnamefont{Wu}},
  \bibinfo{author}{\bibfnamefont{H.}~\bibnamefont{Mei}}, \bibnamefont{and}
  \bibinfo{author}{\bibfnamefont{J.}~\bibnamefont{Yao}},
  \bibinfo{journal}{Science China Physics, Mechanics $\&$ Astronomy}
  \textbf{\bibinfo{volume}{66}}, \bibinfo{pages}{252011}
  (\bibinfo{year}{2023}), ISSN \bibinfo{issn}{1674-7348, 1869-1927},
  \urlprefix\url{https://link.springer.com/10.1007/s11433-022-2045-x}.

\bibitem[{\citenamefont{Xue et~al.}(2023)\citenamefont{Xue, Chen, Chen, Luo,
  Schulze, and Zhou}}]{Xue2023PRC107.044317}
\bibinfo{author}{\bibfnamefont{H.-T.} \bibnamefont{Xue}},
  \bibinfo{author}{\bibfnamefont{Y.-F.} \bibnamefont{Chen}},
  \bibinfo{author}{\bibfnamefont{Q.~B.} \bibnamefont{Chen}},
  \bibinfo{author}{\bibfnamefont{Y.~A.} \bibnamefont{Luo}},
  \bibinfo{author}{\bibfnamefont{H.-J.} \bibnamefont{Schulze}},
  \bibnamefont{and} \bibinfo{author}{\bibfnamefont{X.-R.} \bibnamefont{Zhou}},
  \bibinfo{journal}{Phys. Rev. C} \textbf{\bibinfo{volume}{107}},
  \bibinfo{pages}{044317} (\bibinfo{year}{2023}),
  \urlprefix\url{https://link.aps.org/doi/10.1103/PhysRevC.107.044317}.

\bibitem[{\citenamefont{Tu and Zhou}(2022)}]{Tu2022APJ925.16}
\bibinfo{author}{\bibfnamefont{Z.-H.} \bibnamefont{Tu}} \bibnamefont{and}
  \bibinfo{author}{\bibfnamefont{S.-G.} \bibnamefont{Zhou}},
  \bibinfo{journal}{The Astrophysical Journal} \textbf{\bibinfo{volume}{925}},
  \bibinfo{pages}{16} (\bibinfo{year}{2022}),
  \urlprefix\url{https://dx.doi.org/10.3847/1538-4357/ac3996}.

\bibitem[{\citenamefont{Ren et~al.}(2017)\citenamefont{Ren, Sun, and
  Zhang}}]{Ren2017PRC95.054318}
\bibinfo{author}{\bibfnamefont{S.-H.} \bibnamefont{Ren}},
  \bibinfo{author}{\bibfnamefont{T.-T.} \bibnamefont{Sun}}, \bibnamefont{and}
  \bibinfo{author}{\bibfnamefont{W.}~\bibnamefont{Zhang}},
  \bibinfo{journal}{Phys. Rev. C} \textbf{\bibinfo{volume}{95}},
  \bibinfo{pages}{054318} (\bibinfo{year}{2017}),
  \urlprefix\url{https://link.aps.org/doi/10.1103/PhysRevC.95.054318}.

\bibitem[{\citenamefont{Jennings}(1990)}]{Jennings1990PLB246.1990325}
\bibinfo{author}{\bibfnamefont{B.}~\bibnamefont{Jennings}},
  \bibinfo{journal}{Physics Letters B} \textbf{\bibinfo{volume}{246}},
  \bibinfo{pages}{325} (\bibinfo{year}{1990}), ISSN \bibinfo{issn}{0370-2693},
  \urlprefix\url{https://www.sciencedirect.com/science/article/pii/0370269390906078}.

\bibitem[{\citenamefont{Berger et~al.}(1984)\citenamefont{Berger, Girod, and
  Gogny}}]{Berger1984NPA428.23}
\bibinfo{author}{\bibfnamefont{J.~F.} \bibnamefont{Berger}},
  \bibinfo{author}{\bibfnamefont{M.}~\bibnamefont{Girod}}, \bibnamefont{and}
  \bibinfo{author}{\bibfnamefont{D.}~\bibnamefont{Gogny}},
  \bibinfo{journal}{Nuclear Physics A} \textbf{\bibinfo{volume}{428}},
  \bibinfo{pages}{23} (\bibinfo{year}{1984}), ISSN \bibinfo{issn}{0375-9474},
  \urlprefix\url{https://www.sciencedirect.com/science/article/pii/0375947484902409}.

\bibitem[{\citenamefont{Meng}(1998)}]{Meng1998NPA.635.3}
\bibinfo{author}{\bibfnamefont{J.}~\bibnamefont{Meng}},
  \bibinfo{journal}{Nuclear Physics A} \textbf{\bibinfo{volume}{635}},
  \bibinfo{pages}{3} (\bibinfo{year}{1998}), ISSN \bibinfo{issn}{0375-9474},
  \urlprefix\url{https://www.sciencedirect.com/science/article/pii/S037594749800178X}.

\bibitem[{\citenamefont{Long et~al.}(2010)\citenamefont{Long, Ring, Giai, and
  Meng}}]{Long2010PRC81.024308}
\bibinfo{author}{\bibfnamefont{W.~H.} \bibnamefont{Long}},
  \bibinfo{author}{\bibfnamefont{P.}~\bibnamefont{Ring}},
  \bibinfo{author}{\bibfnamefont{N.~V.} \bibnamefont{Giai}}, \bibnamefont{and}
  \bibinfo{author}{\bibfnamefont{J.}~\bibnamefont{Meng}},
  \bibinfo{journal}{Phys. Rev. C} \textbf{\bibinfo{volume}{81}},
  \bibinfo{pages}{024308} (\bibinfo{year}{2010}),
  \urlprefix\url{https://link.aps.org/doi/10.1103/PhysRevC.81.024308}.

\bibitem[{\citenamefont{Geng et~al.}(2020)\citenamefont{Geng, Xiang, Sun, and
  Long}}]{Geng2020PRC101.064302}
\bibinfo{author}{\bibfnamefont{J.}~\bibnamefont{Geng}},
  \bibinfo{author}{\bibfnamefont{J.}~\bibnamefont{Xiang}},
  \bibinfo{author}{\bibfnamefont{B.~Y.} \bibnamefont{Sun}}, \bibnamefont{and}
  \bibinfo{author}{\bibfnamefont{W.~H.} \bibnamefont{Long}},
  \bibinfo{journal}{Phys. Rev. C} \textbf{\bibinfo{volume}{101}},
  \bibinfo{pages}{064302} (\bibinfo{year}{2020}),
  \urlprefix\url{https://link.aps.org/doi/10.1103/PhysRevC.101.064302}.

\bibitem[{\citenamefont{Geng and Long}(2022)}]{Geng2022PRC105.034329}
\bibinfo{author}{\bibfnamefont{J.}~\bibnamefont{Geng}} \bibnamefont{and}
  \bibinfo{author}{\bibfnamefont{W.~H.} \bibnamefont{Long}},
  \bibinfo{journal}{Phys. Rev. C} \textbf{\bibinfo{volume}{105}},
  \bibinfo{pages}{034329} (\bibinfo{year}{2022}),
  \urlprefix\url{https://link.aps.org/doi/10.1103/PhysRevC.105.034329}.

\bibitem[{\citenamefont{Dover and Gal}(1984)}]{Dover1984PPNP12.171}
\bibinfo{author}{\bibfnamefont{C.~B.} \bibnamefont{Dover}} \bibnamefont{and}
  \bibinfo{author}{\bibfnamefont{A.}~\bibnamefont{Gal}},
  \bibinfo{journal}{Progress in Particle and Nuclear Physics}
  \textbf{\bibinfo{volume}{12}}, \bibinfo{pages}{171} (\bibinfo{year}{1984}),
  ISSN \bibinfo{issn}{0146-6410},
  \urlprefix\url{https://www.sciencedirect.com/science/article/pii/0146641084900048}.

\bibitem[{\citenamefont{and and and}(2013)}]{Wang2013Com.Theor.Phys.60.479}
\bibinfo{author}{\bibnamefont{and}} \bibnamefont{and}
  \bibinfo{author}{\bibnamefont{and}}, \bibinfo{journal}{Communications in
  Theoretical Physics} \textbf{\bibinfo{volume}{60}}, \bibinfo{pages}{479}
  (\bibinfo{year}{2013}),
  \urlprefix\url{https://dx.doi.org/10.1088/0253-6102/60/4/16}.

\bibitem[{\citenamefont{Liu et~al.}(2020)\citenamefont{Liu, Niu, and
  Long}}]{Liu2020PLB806.135524}
\bibinfo{author}{\bibfnamefont{J.}~\bibnamefont{Liu}},
  \bibinfo{author}{\bibfnamefont{Y.~F.} \bibnamefont{Niu}}, \bibnamefont{and}
  \bibinfo{author}{\bibfnamefont{W.~H.} \bibnamefont{Long}},
  \bibinfo{journal}{Physics Letters B} \textbf{\bibinfo{volume}{806}},
  \bibinfo{pages}{135524} (\bibinfo{year}{2020}), ISSN
  \bibinfo{issn}{0370-2693},
  \urlprefix\url{https://www.sciencedirect.com/science/article/pii/S0370269320303282}.

\bibitem[{\citenamefont{Yang et~al.}(2021)\citenamefont{Yang, Sun, Geng, Sun,
  and Long}}]{Yang2021PRC103.014304}
\bibinfo{author}{\bibfnamefont{S.}~\bibnamefont{Yang}},
  \bibinfo{author}{\bibfnamefont{X.~D.} \bibnamefont{Sun}},
  \bibinfo{author}{\bibfnamefont{J.}~\bibnamefont{Geng}},
  \bibinfo{author}{\bibfnamefont{B.~Y.} \bibnamefont{Sun}}, \bibnamefont{and}
  \bibinfo{author}{\bibfnamefont{W.~H.} \bibnamefont{Long}},
  \bibinfo{journal}{Phys. Rev. C} \textbf{\bibinfo{volume}{103}},
  \bibinfo{pages}{014304} (\bibinfo{year}{2021}),
  \urlprefix\url{https://link.aps.org/doi/10.1103/PhysRevC.103.014304}.

\bibitem[{\citenamefont{Wang et~al.}(2021)\citenamefont{Wang, Huang, Kondev,
  Audi, and Naimi}}]{Wang2021CPC45.030003}
\bibinfo{author}{\bibfnamefont{M.}~\bibnamefont{Wang}},
  \bibinfo{author}{\bibfnamefont{W.}~\bibnamefont{Huang}},
  \bibinfo{author}{\bibfnamefont{F.}~\bibnamefont{Kondev}},
  \bibinfo{author}{\bibfnamefont{G.}~\bibnamefont{Audi}}, \bibnamefont{and}
  \bibinfo{author}{\bibfnamefont{S.}~\bibnamefont{Naimi}},
  \bibinfo{journal}{Chinese Physics C} \textbf{\bibinfo{volume}{45}},
  \bibinfo{pages}{030003} (\bibinfo{year}{2021}),
  \urlprefix\url{https://doi.org/10.1088/1674-1137/abddaf}.

\bibitem[{\citenamefont{Zhang et~al.}(2022{\natexlab{c}})\citenamefont{Zhang,
  Cheoun, Choi, Chong, Dong, Dong, Du, Geng, Ha, He
  et~al.}}]{Zhang2022ADNDT144.101488}
\bibinfo{author}{\bibfnamefont{K.}~\bibnamefont{Zhang}},
  \bibinfo{author}{\bibfnamefont{M.-K.} \bibnamefont{Cheoun}},
  \bibinfo{author}{\bibfnamefont{Y.-B.} \bibnamefont{Choi}},
  \bibinfo{author}{\bibfnamefont{P.~S.} \bibnamefont{Chong}},
  \bibinfo{author}{\bibfnamefont{J.}~\bibnamefont{Dong}},
  \bibinfo{author}{\bibfnamefont{Z.}~\bibnamefont{Dong}},
  \bibinfo{author}{\bibfnamefont{X.}~\bibnamefont{Du}},
  \bibinfo{author}{\bibfnamefont{L.}~\bibnamefont{Geng}},
  \bibinfo{author}{\bibfnamefont{E.}~\bibnamefont{Ha}},
  \bibinfo{author}{\bibfnamefont{X.-T.} \bibnamefont{He}},
  \bibnamefont{et~al.}, \bibinfo{journal}{Atomic Data and Nuclear Data Tables}
  \textbf{\bibinfo{volume}{144}}, \bibinfo{pages}{101488}
  (\bibinfo{year}{2022}{\natexlab{c}}), ISSN \bibinfo{issn}{0092-640X},
  \urlprefix\url{https://www.sciencedirect.com/science/article/pii/S0092640X22000018}.

\bibitem[{\citenamefont{Kaur et~al.}(2022)\citenamefont{Kaur, Kanungo,
  Horiuchi, Hagen, Holt, Hu, Miyagi, Suzuki, Ameil, Atkinson
  et~al.}}]{Kaur2022PRL129.142502}
\bibinfo{author}{\bibfnamefont{S.}~\bibnamefont{Kaur}},
  \bibinfo{author}{\bibfnamefont{R.}~\bibnamefont{Kanungo}},
  \bibinfo{author}{\bibfnamefont{W.}~\bibnamefont{Horiuchi}},
  \bibinfo{author}{\bibfnamefont{G.}~\bibnamefont{Hagen}},
  \bibinfo{author}{\bibfnamefont{J.~D.} \bibnamefont{Holt}},
  \bibinfo{author}{\bibfnamefont{B.~S.} \bibnamefont{Hu}},
  \bibinfo{author}{\bibfnamefont{T.}~\bibnamefont{Miyagi}},
  \bibinfo{author}{\bibfnamefont{T.}~\bibnamefont{Suzuki}},
  \bibinfo{author}{\bibfnamefont{F.}~\bibnamefont{Ameil}},
  \bibinfo{author}{\bibfnamefont{J.}~\bibnamefont{Atkinson}},
  \bibnamefont{et~al.}, \bibinfo{journal}{Phys. Rev. Lett.}
  \textbf{\bibinfo{volume}{129}}, \bibinfo{pages}{142502}
  (\bibinfo{year}{2022}),
  \urlprefix\url{https://link.aps.org/doi/10.1103/PhysRevLett.129.142502}.

\bibitem[{\citenamefont{Angeli and Marinova}(2013)}]{Angeli2013ADNDT99.69}
\bibinfo{author}{\bibfnamefont{I.}~\bibnamefont{Angeli}} \bibnamefont{and}
  \bibinfo{author}{\bibfnamefont{K.}~\bibnamefont{Marinova}},
  \bibinfo{journal}{Atomic Data and Nuclear Data Tables}
  \textbf{\bibinfo{volume}{99}}, \bibinfo{pages}{69} (\bibinfo{year}{2013}),
  ISSN \bibinfo{issn}{0092-640X},
  \urlprefix\url{https://www.sciencedirect.com/science/article/pii/S0092640X12000265}.

\bibitem[{\citenamefont{Li et~al.}(2021)\citenamefont{Li, Luo, and
  Wang}}]{Li2021ADNDT140.101440}
\bibinfo{author}{\bibfnamefont{T.}~\bibnamefont{Li}},
  \bibinfo{author}{\bibfnamefont{Y.}~\bibnamefont{Luo}}, \bibnamefont{and}
  \bibinfo{author}{\bibfnamefont{N.}~\bibnamefont{Wang}},
  \bibinfo{journal}{Atomic Data and Nuclear Data Tables}
  \textbf{\bibinfo{volume}{140}}, \bibinfo{pages}{101440}
  (\bibinfo{year}{2021}), ISSN \bibinfo{issn}{0092-640X},
  \urlprefix\url{https://www.sciencedirect.com/science/article/pii/S0092640X21000267}.

\bibitem[{\citenamefont{Sun et~al.}(2008)\citenamefont{Sun, Long, Meng, and
  Lombardo}}]{Sun2008PRC78.065805}
\bibinfo{author}{\bibfnamefont{B.~Y.} \bibnamefont{Sun}},
  \bibinfo{author}{\bibfnamefont{W.~H.} \bibnamefont{Long}},
  \bibinfo{author}{\bibfnamefont{J.}~\bibnamefont{Meng}}, \bibnamefont{and}
  \bibinfo{author}{\bibfnamefont{U.}~\bibnamefont{Lombardo}},
  \bibinfo{journal}{Phys. Rev. C} \textbf{\bibinfo{volume}{78}},
  \bibinfo{pages}{065805} (\bibinfo{year}{2008}),
  \urlprefix\url{https://link.aps.org/doi/10.1103/PhysRevC.78.065805}.

\bibitem[{\citenamefont{Long et~al.}(2012)\citenamefont{Long, Sun, Hagino, and
  Sagawa}}]{Long2012PRC85.025806}
\bibinfo{author}{\bibfnamefont{W.~H.} \bibnamefont{Long}},
  \bibinfo{author}{\bibfnamefont{B.~Y.} \bibnamefont{Sun}},
  \bibinfo{author}{\bibfnamefont{K.}~\bibnamefont{Hagino}}, \bibnamefont{and}
  \bibinfo{author}{\bibfnamefont{H.}~\bibnamefont{Sagawa}},
  \bibinfo{journal}{Phys. Rev. C} \textbf{\bibinfo{volume}{85}},
  \bibinfo{pages}{025806} (\bibinfo{year}{2012}),
  \urlprefix\url{https://link.aps.org/doi/10.1103/PhysRevC.85.025806}.

\bibitem[{\citenamefont{Lv et~al.}(2018)\citenamefont{Lv, Zhang, Zhang, Wu,
  Liu, and Cao}}]{Lv2018CPL35.062102}
\bibinfo{author}{\bibfnamefont{H.}~\bibnamefont{Lv}},
  \bibinfo{author}{\bibfnamefont{S.-S.} \bibnamefont{Zhang}},
  \bibinfo{author}{\bibfnamefont{Z.-H.} \bibnamefont{Zhang}},
  \bibinfo{author}{\bibfnamefont{Y.-Q.} \bibnamefont{Wu}},
  \bibinfo{author}{\bibfnamefont{J.}~\bibnamefont{Liu}}, \bibnamefont{and}
  \bibinfo{author}{\bibfnamefont{L.-G.} \bibnamefont{Cao}},
  \bibinfo{journal}{Chinese Physics Letters} \textbf{\bibinfo{volume}{35}},
  \bibinfo{pages}{062102} (\bibinfo{year}{2018}),
  \urlprefix\url{https://dx.doi.org/10.1088/0256-307X/35/6/062102}.

\end{thebibliography}

\end{document}